\documentclass[journal]{IEEEtran}
\usepackage{amssymb}
\usepackage{amsfonts}
\usepackage{amsmath}
\usepackage{cite}
\usepackage{stfloats}
\usepackage{amsmath}
\usepackage{graphicx}
\usepackage{amsfonts,amsmath,amssymb}
\usepackage{url}
\usepackage{bm}
\usepackage{bbm}
\usepackage{subfigure}
\usepackage{algorithmic}
\usepackage{algorithm}
\usepackage{color}
\newcommand{\Hnull}{\mathcal{H}_0}
\newcommand{\Halt}{\mathcal{H}_1}

\newcommand{\Honull}{\mathcal{{D}}_0}
\newcommand{\Hoalt}{\mathcal{{D}}_1}

\newtheorem{theorem}{\textbf{Theorem}}

\newtheorem{lemma}{\textbf{Lemma}}

\usepackage{graphicx,subfigure,amsmath,amssymb,cite}
\usepackage{float}
\normalsize
\ifCLASSINFOpdf
\else
\fi
\begin{document}
\title{UAV-Enabled Covert Wireless Data Collection}
\author{Xiaobo Zhou, \IEEEmembership{Student Member, IEEE,} Shihao Yan, \IEEEmembership{Member, IEEE,} Feng Shu, \IEEEmembership{Member, IEEE,} Riqing Chen, \IEEEmembership{Member, IEEE,} and Jun Li, \IEEEmembership{Senior Member, IEEE}
\thanks{X. Zhou, J. Li, and F. Shu are with the School of Electronic and Optical Engineering, Nanjing University of Science and Technology, Nanjing, 210094, China (e-mails: \{zxb, jun.li, shufeng\}@njust.edu.cn). X. Zhou is also with the School of Fuyang Normal University, Fuyang, China.}
%\thanks{S. Yan is with the School of Engineering, Macquarie University, Sydney, NSW 2109, Australia (e-mail: shihao.yan@mq.edu.au).}
\thanks{S. Yan is with the School of Engineering, Macquarie University, Sydney, NSW 2109, Australia (e-mail: \{shihao.yan, stephen.hanly\}@mq.edu.au).}
\thanks{R. Chen is with the College of Computer and Information Sciences, Fujian Agriculture and Forestry University, Fuzhou, 350002, China (e-mail: riqing.chen@fafu.edu.cn).}
\thanks{Part of this work has been submitted to IEEE GlobeCOM 2019.}}% <-this % stops a space

\maketitle

\begin{abstract}
 This work considers unmanned aerial vehicle (UAV) networks for collecting data covertly from ground users. The full-duplex UAV intends to gather critical information from a scheduled user (SU) through wireless communication and generate artificial noise (AN) with random transmit power in order to ensure a negligible probability of the SU's transmission being detected by
the unscheduled users (USUs). To enhance the system performance, we jointly design the UAV's trajectory and its maximum AN transmit power together with the user scheduling strategy subject to practical constraints, e.g., a covertness constraint, which is explicitly determined by analyzing each USU's detection performance, and a binary constraint induced by user scheduling. The formulated design problem is a mixed-integer non-convex optimization problem, which is challenging to solve directly, but tackled by our developed penalty successive convex approximation (P-SCA) scheme. An efficient UAV trajectory initialization is also presented based on the Successive Hover-and-Fly (SHAF) trajectory, which also serves as a benchmark scheme. Our examination shows the developed P-SCA scheme significantly outperforms the benchmark scheme in terms of achieving a higher max-min average transmission rate from all the SUs to the UAV.
\end{abstract}

\IEEEpeerreviewmaketitle

\section{Introduction}

Unmanned aerial vehicles (UAVs) communications have attracted significant attention in both military and civilian applications, such as search and rescue, cargo delivery, aerial filming and inspection~\cite{Zeng2016Wireless}.
Different from the traditional terrestrial wireless communications, UAV-enabled wireless communications possess many advantages, such as on-demand and swift deployment, higher network flexibility with the controllable UAV movement, and high possibilities of line-of-sight (LoS) communication links between the UAV and ground users. In particular, the favorable LoS air-to-ground communication links can be efficiently exploited in various UAV-enabled wireless networks for performance enhancement by properly designing the UAV's flight trajectory (e.g., \cite{Zhang2018Joint,Wu2018Capacity,Wu2018Common,Zhan2018Energy}).
However, the LoS air-to-ground communication links also cause UAV communications to suffer from more stringent security issues than the conventional terrestrial wireless communications, since the confidential information transmitted by a UAV is more vulnerable to malicious users when the UAV is in sight.

Recently, several works addressed the wireless communication security of UAV networks from the perspective of physical layer security (e.g., \cite{Zhang2019Securing,Li2018Enabled,Zhou2018Improving,Zhou2019UAV,Cai2018Dual,Wang2017Improving}).
In \cite{Zhang2019Securing}, the authors designed the UAV trajectory and transmit power to enhance the quality of the desired communication link and degrade the eavesdropping link in order to prevent the confidential information from being intercepted by eavesdroppers. Meanwhile, the use of a UAV
as a friendly jammer to assist the terrestrial wireless communication security was considered in \cite{Zhou2018Improving,Li2018Enabled}. Along this direction, the authors of \cite{Zhou2019UAV,Cai2018Dual} considered dual UAV-enabled wireless communications, where one UAV as a transmitter sends confidential information to intended users and the other UAV acting as a jammer generates artificial noise (AN) to create interference to eavesdroppers. It was shown that the communication security of such UAV networks can be enhanced by jointly optimizing the UAV's trajectory and transmit power of the two cooperative UAVs. Furthermore, the work \cite{Wang2017Improving} optimized the location of a UAV (acting as a mobile relay) and its transmit power to improve the security performance of UAV relay networks.

%the secure communication in UAV relay network was considered in \cite{Wang2017Improving}, and the security can be achieved by optimizing the location of mobile relay.

The aforementioned physical layer security technology only addresses protecting the contents of wireless communications in UAV networks. We note that, in some practical scenarios hiding the transmission behavior of a transmitter is explicitly required (e.g., \cite{Zhou2018Joint,Yan2019Hiding,yan2019Low}), which is also desirable in some UAV networks. We note that, once the transmission behavior of a transmitter is detected by malicious users, its location information is exposed, which makes it vulnerable to physical or ongoing attacks. Fortunately, the emerging covert communication technology can hide the very existence of a wireless transmission, i.e., avoiding a wireless transmission being detected by a warden~(e.g., \cite{Bash2013Limits,Shihao2018Delay,Shu2019Delay,Shahzad2018Achieving,goeckel2016covert,BiaoHe2017on,HeB2018Covert,Hu2018Covert,Shahzad2018Relaying,Shahzad2019Covert,Zheng2019Multi,yan2018gaussian}).

Covert communications in additive white Gaussian noise (AWGN) channels was considered in
\cite{Bash2013Limits}, where the authors proved that the transmitter can covertly and reliably transmit no more than $\mathcal{O}(\sqrt{n})$ bits to a receiver. The impact of a finite number of channel uses on covert communication was considered in \cite{Shihao2018Delay}, in which the optimal number of channel uses was derived. Meanwhile, covert communications in full-duplex (FD) networks were examined in the literature (e.g., \cite{Shu2019Delay,Shahzad2018Achieving}), where the transmitter intends to communicate with a FD receiver covertly with the aid of AN transmitted by the receiver. In addition to AN, the impact of noise uncertainty on covert communications was examined in \cite{goeckel2016covert}, where the authors proved that the transmitter can transmit $\mathcal{O}(n)$ bits to receiver covertly and reliably. On this basis, the authors in \cite{BiaoHe2017on} derived the average covert probability and the covert outage probability when a warden's noise power suffers from bounded and unbounded uncertainties. Furthermore, covert communications with poisson field random interferers and covert communications
in relay networks were investigated in \cite{HeB2018Covert} and \cite{Hu2018Covert,Shahzad2018Relaying}, respectively.
Most recently, covert communications with backscatter radio and multi-antenna technology were investigated in \cite{Shahzad2019Covert} and \cite{Zheng2019Multi}, respectively.

\begin{figure}[!t]
  \centering
  % Requires \usepackage{graphicx}
  \includegraphics[width=2.8in]{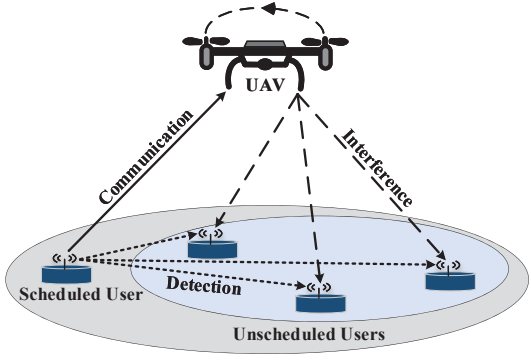}\\
  \caption{Covert communications in the context of UAV data collection networks.}\label{Sys_Sch}
\end{figure}

Data collection is an important application and research topic in the context of Internet of Things (IoT)~\cite{Dong2014UAV}. In conventional IoT scenarios, a sensor node normally has to send its sensing information to a sink node via multi-hop communications, which costs a large amount of energy consumption at the sensor nodes~\cite{Gong2018Flight}. In addition, in some special application scenarios (e.g., remote mountainous or volcanic areas), it is difficult or even impossible to collect information data from all the sensor or sink nodes to the internet. Utilizing a UAV as a data collector, each sensor node can directly transmit its collected information to the UAV and the UAV can sequentially schedule the sensor nodes to collect data from them when it moves sufficiently close to them.
Thus, the use of a UAV as a mobile data collector is highly appealing for saving the energy and proving reliable data collection, which is an significantly important application scenario of UAV networks.
In UAV data collection networks, the time division multiple access (TDMA) protocol is general adopted to save the energy consumption of each sensor node, i.e., the unscheduled sensor nodes can remain in the sleep mode until they receive the waking up beacon signal~\cite{Zhan2018Energy}.
In such data collection scenarios, the sensor nodes may prefer to preserve their privacy (e.g., location information) from each other while transmitting critical information to the UAV, when, for example, the sensor nodes are spies and prefer to hide from each other. In this work, we address this problem and design a UAV-enabled system based on covert wireless communications to enable the sensor nodes to hide their transmissions from each other while conveying critical information to the UAV. In our considered system as shown in Fig.~\ref{Sys_Sch}, the UAV, working on the FD mode, not only collects data from the scheduled user (SU), but also generates AN with random transmit power to create uncertainty at the unscheduled users (USUs) in order to maintain a certain level of covertness. The main contributions of this work are summarized as below.
\begin{itemize}
\item For the first time, we consider UAV-enabled covert data collection based on covert communication techniques to achieve a high-level security and privacy of each ground user in UAV networks. Specifically, we first derive the transmission outage probability from a SU to the UAV, which determines the transmission rate expression based on an outage constraint. We then analyze the detection performance at each USU who serves as the detector warden, i.e., we derive the expressions of the false alarm and miss detection rates, based on which we analytically determine the optimal detection threshold and the corresponding minimum detection error rate. This detection performance analysis enables us to determine the covertness constraint explicitly, which is a main constraint in the system design.

\item In order to enhance the covert data collection performance, we formulate an optimization problem to jointly design the UAV's trajectory, the UAV's maximum AN transmit power, and the user scheduling strategy. This design aims to maximize the minimum average transmission rate (ATR) from all the SUs to the UAV, subject to a covertness constraint, a binary constraint, a transmit power constraint, and the UAV's mobility constraint. The formulated optimization problem is challenging to solve directly, since it is a mixed-integer optimization problem and the optimization variables are closely coupled with each. To tackle it, we develop a penalty successive convex approximation (P-SCA) scheme. Specifically, we first add the penalty term for violating the binary constraint to the objective function and then we apply the first-order restrictive approximation to transform the optimization problem into a convex form, which can be solved with the aid of successive convex approximation (SCA) techniques iteratively.

\item  To improve the convergence rate of the developed P-SCA scheme and achieve a superior covertness performance, we propose an efficient UAV trajectory initialization scheme based on the Successive Hover-and-Fly (SHAF) trajectory, which also serves as a benchmark scheme in this work.
    Our examination shows the developed P-SCA scheme achieves a significantly higher max-min ATR than the benchmark scheme, which demonstrates the necessity of the conducted joint design of the UAV's trajectory and other system parameters.
    Interestingly, our results also show that, as the covertness constraint becomes stricter, the UAV's trajectory achieved by the P-SCA scheme always shrinks inward relative to the region determined by all ground users and the UAV's maximum AN transmit power is dominated by the distance from the UAV to the strongest detector.
\end{itemize}

The reminder of this work is organized as follows. In Section II, we present the considered system model. In Section III, we analyze the detection performance at the USUs. In Section IV, we develop the P-SCA scheme to jointly design the UAV's trajectory and the maximum AN transmit power as well as the user scheduling to maximize the minimum ATR, where the SHAF trajectory initialization scheme is also presented. Section V provides our numerical results draw useful insights on the system design and Section VI presents our conclusion remarks.

%==================================================================================================
\section{System Model}

\subsection{Considered Scenario and Adopted Assumptions}
As shown in Fig.~\ref{Sys_Sch}, in this work we consider covert communications in a UAV network, where a UAV working in the FD mode acts as a mobile data collector to gather information from $K$ users on the ground.
We assume that at most one ground user is scheduled for data transmission at one time instant $t$. The SU (i.e., scheduled user) intends to transmit information to the UAV covertly and does not wish this transmission to be detected by the USUs (i.e., unscheduled users) in order to preserve the privacy of the SU (e.g., hiding the location information of the SU from USUs). In this work, we consider that the UAV is equipped with a receive antenna and a transmit antenna, in which the receive antenna is used for data collection and the transmit antenna is used to assist the covert transmission from the SU to the UAV by generating AN.
The UAV's flight period is set to a finite value $T$ due to the limited battery capacity. During the flight period $T$, the UAV flies at a fixed altitude $H$, which should be properly selected to avoid obstacles. The UAV's trajectory projected onto the horizontal plane is denoted as $\{\mathbf{q}_u(t)\in\mathbb{R}^{2\times 1}, 0\leq t\leq T\}$, while the horizontal coordinate of the $k$-th ground user is denoted by $\mathbf{w}_k\in\mathbb{R}^{2\times 1}$, $k\in\mathcal{K}\triangleq\{1,2,\cdots, K\}$.
To facilitate the UAV trajectory design, we divide the flight period $T$ into $N$ equal-time slots, i.e., $T=N\delta_t$, where $\delta_t$ is the duration of each time slot and should be chosen properly to balance the approximation accuracy and computational complexity. Thus, the UAV's trajectory $\mathbf{q}_u(t)$, $0\leq t\leq T$, can be approximated by $\mathbf{q}_u[n]$, $n\in \mathcal{N}\triangleq\{1,2\cdots,N\}$, where $\mathbf{q}_u[n]=\mathbf{q}_u(n\delta_t)$ denotes the horizontal coordinate of the UAV at the $n$-th time slot.
% to approximate the continuous movement of UAV. The duration of each time slot is given by $\delta_t=\frac{T}{N}$.
% As such, the trajectory of the UAV can be approximated by $\mathbf{q}_u[n]\in\mathbb{R}^{2\times 1}$, $n\in\mathcal{N}\triangleq\{1,2,\cdots,N\}$.
Then, the mobility constraints of the UAV can be written as
\begin{subequations}\label{Mob}
\begin{align}
\mathbf{q}_u[1]&=\mathbf{q}_u[N],\label{Moba}\\
\|\mathbf{q}_u[n+1]-\mathbf{q}_u[n]\|&\leq V_{\max}\delta_t,~n\in \mathcal{N}\setminus\{N\},\label{Mobb}
\end{align}
\end{subequations}
where \eqref{Moba} implies that the UAV has to return the initial location by the end of the last time slot, and \eqref{Mobb} denotes the maximum flight distance during each time slot in which $V_{\max}$ denotes the maximum speed of the UAV.

\subsection{Transmission from the Scheduled User to the UAV}

Following \cite{Zhan2018Energy,Wu2018Common}, we assume that the channels from ground users to UAV are dominated by line-of-sight (LoS). Considering channel reciprocity, at the $n$-th time slot, the channel from the $k$-th ground user to the UAV or the channel from the UAV to the $k$-th ground user is given by~\cite{JointWu2018}
\begin{align}\label{channel}
h_{k,u}[n]=h_{u,k}[n]=\sqrt{\frac{\beta_0}{\|\mathbf{q}_u[n]-\mathbf{w}_k\|^2+H^2}},~\forall k, n,
\end{align}
where $\beta_0$ denotes the channel power gain at a reference distance $1$ meter (m). The channel from $k$-th user to $m$-th user is denoted by $g_{k,m}[n]$, $\forall k, m$, $k\neq m$, and the self-interference channel of UAV is denoted by $g_{u,u}[n]$, $\forall n$, which are subject to quasi-static Rayleigh fading, where $g_{k,m}[n]$ and $g_{u,u}[n]$ follow $\mathcal{CN}(0,\lambda_{k,m})$ and $\mathcal{CN}(0,\lambda_{u,u})$, respectively. In this work, we assume that each ground user only knows the channel distribution information (CDI) between it and other ground users, while the exact instantaneous channel information is unavailable.
In addition, we assume that the location information of all the ground users is known to the UAV, since all these users intend to transmit information to the UAV (i.e., the UAV will collect data from all the ground users). Thus, the UAV knows the channels to all the ground users. Furthermore, we also assume that each ground user knows the channel from itself to the UAV.

%We also assume that each of the USUs possesses complete knowledge of $h_{u,m}[n]$, which is the worst-case scenario for covert communications.

For the $i$-th channel use in the $n$-th time slot, if $k$-th user is scheduled and transmits, the signal received at UAV is given by
\begin{align}\label{y_u}
y_u^{(i)}[n]&=\sqrt{P_k}[n]h_{k,u}[n]s_k(i)+\nonumber\\
&~~~~~~~~~~~~~~~\sqrt{\rho P_u[n]}g_{u,u}[n]s_u(i)+n_u(i),~\forall n,
\end{align}
where $i=1, 2,\cdots, j$ denotes the index of each channel use, $j$ denotes the total number of channel uses in each time slot, $0\leq\rho\leq 1$ denotes the self-interference cancellation coefficient, and $n_u(i)$ is the AWGN at UAV with mean $0$ and variance $\sigma_u^2$, $s_k(i)$ denotes the signal transmitted by the user $k$, following $\mathcal{CN}(0,1)$, and $s_u(i)$ denotes the AN transmitted by the UAV, satisfying $\mathbb{E}[|s_u(i)|^2]=1$. In addition, $P_k[n]$ is the transmit power of the SU $k$. In this work, we assume that $P_k[n]$, $\forall k,n$, is fixed and is publicly known. Furthermore, $P_u[n]$ is the transmit power of AN at the UAV and follows a uniform distribution over the interval $[0, P_{u,\max}[n]]$, where $P_{u,\max}[n]$ is the maximum transmit power of the AN. Specifically, the probability density function (pdf) of $P_u[n]$ is given by
\begin{align}\label{Pu_pdf}
f_{P_u[n]}(x)=
\begin{cases}
\frac{1}{P_{u,\max}[n]},&0\leq x\leq P_{u,\max}[n],\\
0,&\mathrm{otherwise}.
\end{cases}
\end{align}
We assume that the USUs only know the distribution information of the UAV's AN transmit power.
We note that introducing the randomness of the AN transmit power is to create an uncertainty of the received power at the USUs to assist the SU's covert transmission.

We use $x_k[n]$ to denote the scheduling variable, where $x_k[n]=1$ if ground user $k$ is scheduled at time slot $n$, and $x_k[n]=0$ otherwise. In addition, we note that at most one ground user is scheduled by the UAV at each time slot. As such, we have the following constraint
\begin{align}\label{schedule}
\sum_{k=1}^K x_k[n]\leq 1,\forall n, ~~x_k[n]\in\{0,1\},\forall k, n.
\end{align}

Following \eqref{y_u}, if user $k$ is scheduled for communication at time slot $n$, the channel capacity from this user to the UAV is given by
\begin{align}\label{Rate1}
C_k[n]=\log_2\left(1+\frac{P_k[n]|h_{k,u}[n]|^2}{\rho P_u[n]|g_{u,u}[n]|^2+\sigma_u^2}\right).
\end{align}
We note that $P_u[n]$ is controlled by the UAV and thus it is known to the UAV. The transmission from the user $k$ to the UAV can still suffer outage due to the random self-interference channel $g_{u,u}[n]$. The transmission outage probability between user $k$ and UAV is given by
%Since the randomness of $g_{u,u}[n]$ and $P_u[n]$, we consider the average performance of the achievable rate in each time slot, which is given by $\mathbb{E}[R_k[n]]$. In addition, we observe that $R_k[n]$ is a convex function with respect to $P_u[n]|g_{u,u}[n]|^2$. As such, applying the Jensen's inequality, we have
\begin{align}\label{Outage}
&\mathrm{Pr}\{C_k[n] < R_k[n]\}\nonumber\\
&=\mathrm{Pr}\left\{|g_{u,u}[n]|^2>\frac{P_k[n]|h_{k,u}[n]|^2}{\rho P_u[n](2^{R_k[n]}-1)}-\frac{\sigma_u^2}{\rho P_u[n]}\right\}\nonumber\\
%&=\exp{\left[\frac{-1}{\lambda_{u,u}}\left(\frac{P_k[n]|h_{k,u}[n]|^2}{\rho P_u[n](2^{R_k[n]}-1)}-\frac{\sigma_u^2}{\rho P_u[n]}\right)\right]}\nonumber\\
&=\exp{\left[\frac{-1}{\rho P_u[n]\lambda_{u,u}}\left(\frac{P_k[n]|h_{k,u}[n]|^2}{2^{R_k[n]}-1}-\sigma_u^2\right)\right]},
\end{align}
where $R_k[n]$ is the transmission rate from user $k$ to the UAV at time slot $n$. We note that transmission outage probability is an increasing function of $P_u[n]$. As such, an upper bound on transmission outage probability from user $k$ to the UAV at the $n$-th time slot is given by
%\begin{align}\label{Outage2}
%&\mathrm{Pr}\{C_k[n] < R_k[n]\}\leq p_{k}^{out}[n]
%\end{align}
%where
\begin{align}\label{Outage3}
&p_{k}^{out}[n]=\nonumber\\
&\exp{\left[\frac{-1}{\rho P_{u,\max}[n]\lambda_{u,u}}\left(\frac{P_k[n]|h_{k,u}[n]|^2}{2^{R_k[n]}-1}\!-\!\sigma_u^2\right)\right]},\forall k,n.
\end{align}

In this work, we consider a reliability constraint on the transmission from the user $k$ to the UAV, i.e.,  the upper bound on the transmission outage probability is no more than $\epsilon$, i.e., $p_k^{out}[n]\leq\epsilon$, where $\epsilon$ denotes the maximum tolerable outage probability. From \eqref{Outage3}, we see that $p_k^{out}[n]$ is an increasing function of $R_k[n]$. As such, in order to maximize the transmission rate, $p_k^{out}[n]=\epsilon$ is always guaranteed. Therefore, the transmission rate of user $k$ can be expressed as
\begin{align}\label{Rate2}
R_k[n]\!=\!\log_2\left(1\!+\!\frac{P_k[n]|h_{k,u}[n]|^2}{-\rho P_{u,\max}[n]\lambda_{u,u}\ln{\epsilon}+\sigma_u^2}\right),\forall k,n.
\end{align}
%We observe from \eqref{Outage3} that
%where \eqref{Ave_Rb} is due to the fact that the random variables $P_u[n]$ and $|g_{u,u}[n]|^2$ are independent.
\subsection{Binary Hypothesis Testing at the Unscheduled Users}

We note that, each of the USUs faces a binary hypothesis testing problem, i.e., the USUs have to decide whether the SU transmitted to the UAV. Here, the USUs do not cooperate to conduct the detection, since the ground users are distributed and each of them potentially serves as a SU in some specific time slot.
For the $i$-th channel use in the $n$-th time slot, the received signal at the $m$-th USU from the $k$-th SU is given by
\begin{align}\label{Hy_test}
&y_m^{(i)}[n]=\nonumber\\
&\begin{cases}
\sqrt{P_u[n]}h_{u,m}[n]s_u(i)+n_m(i),&\!\Hnull,\!\\
\!\sqrt{P_k[n]}g_{k,m}[n]s_k(i)\!+\!\sqrt{P_u[n]}h_{u,m}[n]s_u(i)\!+\!n_m(i),\!&\!\Halt,\!
\end{cases}
\end{align}
where $m\in\mathcal{K}\setminus\{k\}$, $n_m(i)$ is the AWGN at the $m$-th USU with mean $0$ and variance $\sigma_m^2$, $\Hnull$ is the null hypothesis where the $k$-th SU did not transmit, while $\Halt$ is the alternative hypothesis where the $k$-th SU did transmit to the UAV.
%We assume that each of USUs uses a radiometer as the detector at each time slot due to its simplicity.
We assume that each of USUs uses a radiometer as the detector at each time slot.
This assumption is justified by the fact that the radiometer is the optimal detector in the considered system model, which can be proved along the same lines as the proof in \cite{Sobers2017Covert,Hu2019Covert}.
%We note that this assumption is justified and the optimality of radiometer can be proved along the same lines as the proof in \cite{Sobers2017Covert}.
Thus, each USU conduct a threshold test on the average power received in time slot $n$, which is given by
\begin{align}\label{Like_Test}
T_m[n]\triangleq \frac{1}{j}\sum_{i=1}^j|y_m^{(i)}[n]|^2\mathop{\gtreqless}\limits_{\Honull}^{\Hoalt} \tau_m[n],~m\in\mathcal{K}\setminus\{k\},
\end{align}
where $T_m[n]$ and $\tau_{m}[n]$ are the average power received at the $m$-th USU and its corresponding detection threshold at the $n$-th time slot, respectively, while $\Honull$ and $\Hoalt$ are the decisions in favor of $\Hnull$ and $\Halt$, respectively. In this work, we adopt a widely used assumption in covert communications \cite{HeB2018Covert,Shahzad2018Achieving,BiaoHe2017on,goeckel2016covert}, which is that $j\rightarrow\infty$. This allows each of USUs to observe an infinite number of samples, which is an upper bound on the number of samples that each USU can receive in practice.
As $j\rightarrow\infty$, $T_m[n]$ can be written as
\begin{align}\label{Tm}
T_m[n]=
\begin{cases}
P_u[n]|h_{u,m}[n]|^2+\sigma_m^2,&\Hnull,\\
P_k[n]|g_{k,m}[n]|^2+P_u[n]|h_{u,m}[n]|^2\!+\!\sigma_m^2,&\Halt.
\end{cases}
\end{align}

Following \eqref{Tm}, at the $n$-th time slot, the false alarm rate and miss detection rate at the $m$-th USU when $k$-th user is scheduled are denoted as $\alpha_m[n]=\mathrm{Pr}\{\Hoalt|\Hnull\}$ and $\beta_{k,m}[n]= \mathrm{Pr}\{\Honull|\Halt\}$, respectively, which are given as
\begin{align}
\alpha_m[n]&=\mathrm{Pr}\left\{P_u[n]|h_{u,m}[n]|^2+\sigma_m^2\geq \tau_m[n]\right\},\label{PF}\\
\!\beta_{k,m}[n]&\!=\!\mathrm{Pr}\big\{\!P_k[n]|g_{k,m}[n]|^2\!\!+\!\!P_u[n]|h_{u,m}[n]|^2\!\!+\!\!\sigma_m^2\!\leq \tau_m[n]\big\},\label{PM}
\end{align}
respectively. Then, the detection error rate for the $m$-th USU's detection at the $n$-th time slot when $k$-th user is scheduled is given by
\begin{align}\label{Tot_Err}
\xi_{k,m}[n]=\alpha_m[n]+\beta_{k,m}[n],~\forall n.
\end{align}

In covert communications, the USUs aim to achieve the minimum detection error rate, denoted by $\xi_{k,m}^\ast[n]$, while the SU tries to ensure this minimum detection error rate at each USU being no less than a specific value, i.e., $\xi_{k,m}^\ast[n]\geq 1-\varepsilon$, $\forall n$, where $\varepsilon$ is an arbitrary small constant to determine the required covertness. In the following section, we first derive the optimal detection threshold for each USU and then determine the minimum detection error rate in order to determine the explicit covertness constraint. In section IV, we design the UAV's trajectory and the maximum AN transmit power $P_{u,\max}[n]$ together with the user scheduling strategy to maximize the minimum ATR among all the ground users subject to the determined covertness constraint.

\section{Detection Performance Analysis at the Unscheduled Users}

In this section, we first derive the explicit expressions for the false alarm rate and miss detection rate, respectively.  Then we derive the optimal detection threshold for each USU and the corresponding minimum detection error rate.

\subsection{False Alarm Rate}

Following \eqref{Pu_pdf} and \eqref{PF}, the false alarm rate $\alpha_m[n]$ for the $m$-th USU at the $n$-th time slot when user $k$ is scheduled, is given by
\begin{align}
\alpha_m[n]&=
\begin{cases}\label{False}
1,&\tau_m[n]\leq \sigma_m^2,\\
1-\frac{\tau_m[n]-\sigma_m^2}{\varrho_{u,m}[n]},&\sigma_m^2<\tau_m[n]\leq \varrho_{u,m}[n]+\sigma_m^2,\\
0,&\tau_m[n]> \varrho_{u,m}[n]+\sigma_m^2,
\end{cases}
\end{align}
where $\varrho_{u,m}[n]\triangleq P_{u,\max}[n]|h_{u,m}[n]|^2$.

\subsection{Miss Detection Rate}

We recall that the USUs only have the CDI of the channels to the SU. As such, the miss detection rate $\beta_{k,m}[n]$, defined in \eqref{PM}, involves two random variables with different distributions, where $P_k[n]|g_{k,m}[n]|^2$ follows an exponential distribution with parameter $\frac{1}{P_k[n]\lambda_{k,m}}$ and $P_u[n]|h_{u,m}[n]|^2$ follows a uniform distribution over the interval $\left[0,P_{u,\max}[n]|h_{u,m}[n]|^2\right]$. As a result, we need to derive the pdf of $Z_{u,k,m}[n]$, which is defined as
\begin{align}\label{def_z}
Z_{u,k,m}[n]\triangleq X_{u,m}[n]+Y_{k,m}[n],
\end{align}
in order to derive the miss detection rate, where $X_{u,m}[n]\triangleq P_u[n]|h_{u,m}[n]|^2$ and $Y_{k,m}[n]\triangleq P_k[n]|g_{k,m}|^2$. To this end, we first present the following lemma.
\begin{lemma}\label{lemma1}
The pdf of the random variable $Z_{u,k,m}[n]$ defined in \eqref{def_z} is given by
\begin{align}
&f_{z_{u,k,m}[n]}(z)=\nonumber\\
&\begin{cases}\label{pdf}
\frac{1-\exp{\left(\frac{-z}{P_k[n]\lambda_{k,m}}\right)}}{\varrho_{u,m}[n]},&0< z\leq \varrho_{u,m}[n],\\
\frac{\!\exp{\big(\frac{z-\varrho_{u,m}[n]}{-P_k[n]\lambda_{k,m}}\big)}\!-\exp{\big(\frac{-z}{\!P_k[n]\lambda_{k,m}\!}\big)}}{\varrho_{u,m}[n]},& z>\varrho_{u,m}[n],\\
0,&z\leq 0.
\end{cases}
\end{align}
\end{lemma}

\begin{IEEEproof}
The pdf of $Z_{u,k,m}[n]$ can be written as
\begin{align}\label{pdf_1}
f_{z_{u,k,m}[n]}(z)=\int_{-\infty}^\infty f_{x_{u,m}[n]}(x)f_{y_{k,m}[n]}(z-x)\mathrm{d}x,
\end{align}
where $f_{x_{u,m}[n]}(x)$ and $f_{y_{k,m}[n]}(y)$ are the pdfs of the random variables $X_{u,m}[n]$ and $Y_{k,m}[n]$, respectively. As such, we have $0 \leq x \leq \varrho_{u,m}[n]$ and $x\leq z\leq \infty$. For $0< z\leq \varrho_{u,m}[n]$, the pdf of $Z_{u,k,m}[n]$ can be written as
\begin{align}\label{pdf_2}
f_{z_{u,k,m}[n]}(z)&=\int_{0}^z \frac{\exp{\left(\frac{z-x}{-P_k[n]\lambda_{k,m}}\right)}}{\varrho_{u,m}[n]P_k[n]\lambda_{k,m}}\mathrm{d}x\nonumber\\
&=\frac{1}{\varrho_{u,m}[n]}\left[1-\exp{\left(\frac{-z}{P_k[n]\lambda_{k,m}}\right)}\right].
\end{align}
For $\varrho_{u,m}[n]< z< \infty$, the pdf of $Z_{u,k,m}[n]$ is given by
\begin{align}\label{pdf_3}
&f_{z_{u,k,m}[n]}(z)=\int_{0}^{\varrho_{u,m}[n]} \frac{\exp{\left(\frac{z-x}{-P_k[n]\lambda_{k,m}}\right)}}{\varrho_{u,m}[n]P_k[n]\lambda_{k,m}}\mathrm{d}x\nonumber\\
&\!=\!\frac{1}{\varrho_{\!u,m}[n]\!}\left[\exp{\left(\frac{\varrho_{u,m}[n]\!-\!z}{P_k[n]\lambda_{k,m}}\right)}\!-\!\exp{\left(\frac{-z}{P_k[n]\lambda_{k,m}}\right)}\right].
\end{align}
In addition, for $Z_{u,k,m}[n]\leq 0$, the pdf of $Z_{u,k,m}[n]$ is given by $f_{z_{u,k,m}[n]}(z)=0$. Combining the results in these three cases leads to the desired result in \eqref{pdf}.
\end{IEEEproof}

In order to derive the miss detection rate for $m$-th USU at the $n$-th time slot when $k$-th user is scheduled, we first rewrite the miss detection rate $\beta_{k,m}[n]$ as
\begin{align}\label{PM2}
\beta_{k,m}[n]=\mathrm{Pr}\big\{Z_{u,k,m}[n]\leq \tau_m[n]-\sigma_m^2\big\}.
\end{align}
Then, following Lemma \ref{lemma1}, the miss detection rate for $m$-th USU at the $n$-th time slot when $k$-th user is scheduled is given by
\begin{align}
\beta_{k,m}[n]&=
\begin{cases}\label{PM3}
0,&\tau_{m}[n]\leq \sigma_m^2,\\
\varsigma_{k,m}[n],&\sigma_m^2<\tau_m[n]\leq \varrho_{u,m}[n]+\sigma_m^2,\\
\phi_{k,m}[n],&\tau_m[n]> \varrho_{u,m}[n]+\sigma_m^2,
\end{cases}
\end{align}
where
\begin{align}
&\varsigma_{k,m}[n]\!=\!\int_{0}^{\tau_m[n]\!-\!\sigma_m^2}\frac{\exp{\left(\frac{z-x}{-P_k[n]\lambda_{k,m}}\right)}}{\varrho_{u,m}[n]P_k[n]\lambda_{k,m}}\mathrm{d}x\nonumber\\
&=\frac{\tau_m[n]\!-\!\sigma_m^2}{\varrho_{u,m}[n]}-\frac{P_k[n]\lambda_{k,m}}{\varrho_{u,m}[n]}
\left[1\!-\!\exp{\left(\frac{\tau_m[n]-\sigma_m^2}{-P_k[n]\lambda_{k,m}}\right)}\right],\label{PM4}\\
&\!\phi_{k,m}[n]\!=\!\int_{0}^{\varrho_{\!u,m}[n]\!} \!f_{\!z_{u,k,m}[n]\!}(z)\mathrm{d}z\!+\!\int_{\!\varrho_{u,m}[n]\!}^{\tau_m[n]-\sigma_m^2} \!f_{\!z_{u,k,m}[n]\!}(z)\mathrm{d}z\!\nonumber\\
&~~~~~~~~=1-\frac{P_k[n]\lambda_{k,m}}{\varrho_{u,m}[n]}\Bigg[\exp{\left(\frac{\tau_m[n]-\sigma_m^2-\varrho_{u,m}[n]}{-P_k[n]\lambda_{k,m}}\right)}\nonumber\\
&~~~~~~~~~~~-\exp{\left(\frac{\tau_m[n]-\sigma_m^2}{-P_k[n]\lambda_{k,m}}\right)}\Bigg].\label{PM5}
\end{align}

\subsection{Optimal Detection Threshold and the Minimum Detection Error Rate}

Following \eqref{False} and \eqref{PM3}, we derive the optimal detection threshold denoted by $\tau_m^*[n]$ and the corresponding minimum detection error rate denoted by $\xi_{k,m}^*[n]$ in the following theorem.
\begin{theorem}\label{theorem1}
The optimal detection threshold for $m$-th USU at the $n$-th time slot when user $k$ is scheduled is given by $\tau_{m}^\ast[n]=\varrho_{u,m}[n]+\sigma_m^2$ and the corresponding minimum detection error rate is given by
\begin{align}\label{ksi1}
&\xi_{k,m}^\ast[n]=1-\frac{P_k[n]\lambda_{k,m}}{\varrho_{u,m}[n]}\left[1-\exp{\left(\frac{-\varrho_{u,m}[n]}{P_k[n]\lambda_{k,m}}\right)}\right].
\end{align}
\end{theorem}

\begin{IEEEproof}
Following \eqref{False} and \eqref{PM3}, the detection error rate for the $m$-th user at the $n$-th time slot is given by
\begin{align}
\!\xi_{k,m}[n]\!\!=\!
&\begin{cases}\label{ksi2}
1,&\tau_{m}[n]\leq \sigma_m^2,\\
1\!-\!\hat{\varsigma}_{k,m}[n],
&\sigma_m^2\!<\!\tau_m[n]\!\leq\! \varrho_{u,m}[n]+\sigma_m^2,\\
\phi_{k,m}[n],&\tau_m[n]> \varrho_{u,m}[n]+\sigma_m^2,
\end{cases}
\end{align}
where %$\hat{\varsigma}_{k,m}[n]\triangleq\frac{P_k[n]\lambda_{k,m}}{\varrho_{u,m}[n]}\left[1-\exp{\left(\frac{\tau_m[n]-\sigma_m^2}{-P_k[n]\lambda_{k,m}}\right)}\right]$.
\begin{align}
\hat{\varsigma}_{k,m}[n]\triangleq\frac{P_k[n]\lambda_{k,m}}{\varrho_{u,m}[n]}\left[1-\exp{\left(\frac{\tau_m[n]-\sigma_m^2}{-P_k[n]\lambda_{k,m}}\right)}\right].
\end{align}
We note that $\xi_{k,m}[n]=1$ is the worst scenario for the $m$-th USU, i.e., its detection performance is the same as that of a random guess. As such, the $m$-th USU will not set its detection threshold as $\tau_{m}[n]\leq \sigma_m^2$. As per \eqref{ksi2}, we note that $\xi_{k,m}[n]$ monotonically decreases with $\tau_{m}[n]$ for $\sigma_m^2<\tau_m[n]\leq \varrho_{u,m}[n]+\sigma_m^2$, while $\xi_{k,m}[n]$ monotonically increases with $\tau_{m}[n]$ for $\tau_m[n]> \varrho_{u,m}[n]+\sigma_m^2$. Considering that $\xi_{k,m}[n]$ is a continuous function of $\tau_{m}[n]$ in \eqref{ksi2}, we conclude that the optimal detection threshold is given by $\tau_{m}^\ast[n]=\varrho_{u,m}[n]+\sigma_m^2$. Substituting $\tau_{m}^*[n]$ into \eqref{ksi2}, we obtain the minimum detection error rate at the $m$-th USU as given in \eqref{ksi1}.
\end{IEEEproof}

We recall that $P_k[n]$ is the transmit power of the SU $k$ and $\varrho_{u,m}[n]= P_{u,\max}[n]|h_{u,m}[n]|^2$, where $P_{u,\max}[n]$ is the maximum transmit power of AN and $h_{u,m}[n]$ is the channel between the UAV and the $m$-th USU. Thus, as per Theorem~\ref{theorem1} we note that this minimum detection error rate $\xi_{k,m}^\ast[n]$ decreases with the transmit power of the SU but increases with the UAV's AN maximum transmit power. In addition, we note that as the quality of the channel $h_{u,m}[n]$ increases, $\xi_{k,m}^\ast[n]$ increases, which indicates that the UAV may prefer to fly close to USUs in order to maintain a high covertness. Following Theorem~\ref{theorem1}, we can explicitly determine the covertness constraint $\xi_{k,m}^\ast[n]\geq 1-\varepsilon$, based on which we will tackle the UAV's covert data collection problem in the following section.

\section{UAV's Covert Data Collection Design}

In this section, we aim to design the user scheduling and the UAV's trajectory as well as the UAV's maximum AN transmit power to ensure that the UAV can collect data from each ground user reliably and covertly. To this end, we first formulate the UAV's optimization problem and then develop an efficient algorithm to solve it.

\subsection{Optimization Problem Formulation}

For ease of presentation, we define $\mathbf{X}=\{x_k[n],\forall k,n\}$, $\mathbf{Q}=\{\mathbf{q}_u[n],\forall n\}$, and $\mathbf{P}_{\mathrm{U}}=\{P_{u,\max}[n],\forall n\}$, where we recall that $x_k[n]$ is the user scheduling variable, $\mathbf{q}_u[n]$ is the UAV trajectory, and $P_{u,\max}[n]$ is the UAV's maximum AN transmit power. In order to ensure that the UAV can collect data from each ground user and guarantee the fairness among all users, our design aim is set to maximize the minimum ATR among all ground users by jointly designing the user scheduling $\mathbf{X}$, the UAV trajectory $\mathbf{Q}$, and the UAV's maximum AN transmit power $\mathbf{P}_{\mathrm{U}}$.
Then, the design optimization problem at the UAV is formulated as
\begin{subequations}\label{PF1}
\begin{align}
&(\mathbf{P1}):~\max_{\mathbf{Q},\mathbf{X},\mathbf{P}_{\mathrm{U}}}\min_k~\frac{1}{N}\sum_{n=1}^N x_k[n]R_k[n]\label{PF1a}\\
%&\mathrm{s.t.}~\sum_{k=1}^K\sum_{m=1,m\neq k}^Kx_k[n]\xi_{k,m}^*[n]\geq 1-\varepsilon,~\forall n,\label{PF1b}\\
&\mathrm{s.t.}~\sum_{k=1}^Kx_k[n]\min_{m\in\mathcal{K}\setminus\{k\}}\xi_{k,m}^*[n]\geq 1-\varepsilon,~\forall n,\label{PF1b}\\
&~~~~~\sum_{k=1}^K x_k[n]\leq 1, ~\forall n,\label{PF1c}\\
&~~~~~x_k[n]\in\{0,1\},~\forall k, n,\label{PF1d}\\
&~~~~~P_{u,\max}[n]\leq P_{\max}^u,~\forall n,\label{PF1e}\\
&~~~~~\mathbf{q}_u[1]=\mathbf{q}_u[N],~\label{PF1f}\\
&~~~~~\|\mathbf{q}_u[n+1]-\mathbf{q}_u[n]\|\leq V_{\max}\delta_t,~n\in\mathcal{N}\setminus\{N\},\label{PF1g}
\end{align}
\end{subequations}
where $R_k[n]$ defined in \eqref{Rate2} denotes the transmission rate of user $k$. The constraint \eqref{PF1b} is to ensure the covertness of the uplink transmission from the SU to the UAV, where $\xi_{k,m}^*[n]$ is defined in \eqref{ksi1} and $\varepsilon$ is an arbitrarily
small value determining the required covertness. In addition, \eqref{PF1c} and \eqref{PF1d} are the user scheduling constraints, which ensure that at most one ground user is scheduled at each time slot.
Furthermore, \eqref{PF1e} denotes the maximum AN transmit power constraint of UAV, while \eqref{PF1f} and \eqref{PF1g} are UAV's mobility constraints.

%We note that the constraint \eqref{PF1c} must hold with equality at any feasible solution, otherwise \eqref{PF1b} may infeasible due to $x_k[n]=0$, $\forall k$.

We note that the constraint \eqref{PF1c}, the power constraint \eqref{PF1e}, and the mobility constraints \eqref{PF1f} and \eqref{PF1g} are convex, while the objective function \eqref{PF1a}, the covertness constraint \eqref{PF1b}, and the user scheduling constraint \eqref{PF1d} are non-convex. Meanwhile, the user scheduling variables $x_k[n]$, $\forall k,n$, are binary. Thus, the optimization problem ($\mathbf{P1}$) is a mixed-integer non-convex optimization problem, which is difficult to be optimally solved~\cite{JointWu2018}. In the following, we develop an algorithm to solve ($\mathbf{P1}$).

\subsection{P-SCA scheme for Solving the Optimization Problem ($\mathbf{P1}$)}
\newcounter{mytempeqncnt1}
\begin{figure*}[tp]
\normalsize
\setcounter{mytempeqncnt1}{\value{equation}}
\setcounter{equation}{38}
\begin{align}\label{lem2_2}
&R^{lo}_k\left(\mathbf{q}_u[n],P_{u,\max}[n],\mathbf{\tilde{q}}_u[n],\tilde{P}_{u,\max}[n]\right)\triangleq \nonumber\\ &\log_2\left(1+\frac{\frac{\beta_0P_k[n]}{\|\mathbf{\tilde{q}}_u[n]-\mathbf{w}_k\|^2+H^2}}{-\rho \tilde{P}_{u,\max}[n]\lambda_{u,u}\ln{\epsilon}+\sigma_u^2}\right)+ \frac{\frac{-\beta_0P_k[n]\left(\|\mathbf{q}_u[n]-\mathbf{w}_k\|^2-\|\mathbf{\tilde{q}}_u[n]-\mathbf{w}_k\|^2\right)}{\left(\|\mathbf{\tilde{q}}_u[n]-\mathbf{w}_k\|^2+H^2\right)\ln2}+\frac{\beta_0\rho\lambda_{u,u}\ln{\epsilon}P_k[n](P_{u,\max}[n]-\tilde{P}_{u,\max}[n])}{\left(-\rho \tilde{P}_{u,\max}[n]\lambda_{u,u}\ln{\epsilon}+\sigma_u^2\right)\ln2}}{\left(-\rho \tilde{P}_{u,\max}[n]\lambda_{u,u}\ln{\epsilon}+\sigma_u^2\right)\left(\|\mathbf{\tilde{q}}_u[n]-\mathbf{w}_k\|^2+H^2\right)+\beta_0P_k[n]}.
\end{align}
\setcounter{equation}{\value{mytempeqncnt1}}
\hrulefill
\vspace*{4pt}
\end{figure*}
In this subsection, we develop a P-SCA optimization framework to tackle the formulated mixed-integer optimization problem ($\mathbf{P1}$). The central idea of the P-SCA scheme is to first add the penalty term that violates the binary constraint to the objective function and then apply the SCA technique to solve the resultant optimization problem iteratively.

To proceed, we first equivalently rewrite the binary constraint \eqref{PF1d} into continues constraints, which are given by
\begin{subequations}\label{PS1}
\begin{align}
x_k[n]-x_k[n]^2\leq 0,~\forall k, n,\label{PS1a}\\
0\leq x_k[n]\leq 1,~\forall k, n.\label{PS1b}
\end{align}
\end{subequations}
We note that \eqref{PS1a} implies that $x_k[n]\leq 0$ or $x_k[n]\geq 1$ must hold. We jointly consider \eqref{PS1a} and \eqref{PS1b}, leading to  $x_k[n]=0$ or $x_k[n]=1$, which is equivalent the result ensured by the original binary constraint \eqref{PF1d}. In fact, \eqref{PS1} can be further simplified as
\begin{subequations}\label{PS2}
\begin{align}
\sum_{n=1}^N\sum_{k=1}^K\left(x_k[n]-x_k[n]^2\right)\leq 0,~\label{PS2a}\\
0\leq x_k[n]\leq 1,~\forall k, n.\label{PS2b}
\end{align}
\end{subequations}
Following the above transformation and introducing a slack variable $\eta$, the optimization problem ($\mathbf{P1}$) can be equivalently rewritten as
\begin{subequations}\label{PF2_1}
\begin{align}
&(\mathbf{P2.1}):~
\max_{\eta,\mathbf{Q},\mathbf{X},\mathbf{P}_{\mathrm{U}}}~\eta\label{PF2_1a}\\
%&\mathrm{s.t.}~\sum_{k=1}^K\sum_{m=1,m\neq k}^Kx_k[n]\xi_{k,m}^*[n]\geq 1-\varepsilon,~\forall n,\label{PF1b}\\
&\mathrm{s.t.}~\frac{1}{N}\sum_{n=1}^N x_k[n]\log_2\left(1+\frac{\frac{\beta_0 P_k[n]}{\|\mathbf{q}_u[n]-\mathbf{w}_k\|^2+H^2}}{-\rho P_{u,\max}[n]\lambda_{u,u}[n]\ln{\epsilon}+\sigma_u^2}\right)\nonumber\\
&~~~~~~~~~~~~~~~~~~~~~~~~~~~~~~~~~~~~~~~~~~~~~~~~\geq \eta, ~\forall k,\label{PF2_1b}\\
&~~~~~\sum_{k=1}^Kx_k[n]\min_{m\in\mathcal{K}\setminus\{k\}}\left(1-\bar{\xi}_{k,m}^*[n]\right)\geq 1-\varepsilon,~\forall n,\label{PF2_1c}\\
&~~~~~\eqref{PF1c},\eqref{PF1e},\eqref{PF1f},\eqref{PF1g},\eqref{PS2a},\eqref{PS2b},\notag
\end{align}
\end{subequations}
where $\bar{\xi}_{k,m}^*[n]$ in \eqref{PF2_1c} is defined as
\begin{align}\label{PS3}
&\bar{\xi}_{k,m}^*[n]=\frac{\|\mathbf{q}_u[n]-\mathbf{w}_m\|^2+H^2}{\bar{\beta}_{k,m}[n]P_{u,\max}[n]}\times\nonumber\\
&~~~~~~~~~~~~~~~~~~\left[1\!-\!\exp{\left(\frac{-\bar{\beta}_{k,m}[n]P_{u,\max}[n]}{\|\mathbf{q}_u[n]-\mathbf{w}_m\|^2+H^2}\right)}\right],
\end{align}
and $\bar{\beta}_{k,m}[n]\triangleq\frac{\beta_0}{P_k[n]\lambda_{k,m}}$.
Although we have transformed the mixed-integer optimization problem ($\mathbf{P1}$) into the continues optimization problem ($\mathbf{P2.1}$), it is still difficult to tackle due to the joint existence of the constraints \eqref{PS2a} and \eqref{PS2b} as well as the non-convex constraints \eqref{PF2_1b} and \eqref{PF2_1c}. In general, we can apply the first-order restrictive approximation to transform a non-convex constraint set into a convex set, and then employ the SCA technique to solve the optimization problem iteratively. Unfortunately, direct applying SCA technique to ($\mathbf{P2.1}$) will lead to the fact that the feasible solutions to the optimization problem are not achievable  due to the constraints \eqref{PS2a} and \eqref{PS2b}~\cite{Vu2018Weighted}.
Inspired by the recent works \cite{Lipp2016,Vu2016Max}, we overcome this issue by using the penalty method to convert the constraint \eqref{PS2a} into the objective function. Then, the optimization problem ($\mathbf{P2.1}$) can be rewritten as
\begin{subequations}\label{PF2_2}
\begin{align}
&(\mathbf{P2.2}):~
\max_{\eta,\mathbf{Q},\mathbf{X},\mathbf{P}_{\mathrm{U}},\phi}~\eta-\mu\phi\label{PF2_2a}\\
&\mathrm{s.t.}~\sum_{n=1}^N\sum_{k=1}^K\left(x_k[n]-x_k[n]^2\right)\leq \phi,\label{PF2_2b}\\
&~~~~~\eqref{PF1c},\eqref{PF1e},\eqref{PF1f},\eqref{PF1g},\eqref{PS2b},\eqref{PF2_1b},\eqref{PF2_1c}, \notag
\end{align}
\end{subequations}
where $\mu>0$ is a given penalty parameter and $\phi$ is an introducing slack variable. We note that introducing the slack variable $\phi$ can extend the feasible set of ($\mathbf{P2.2}$) relative to that of ($\mathbf{P2.1}$). Initially, we choose a small value of $\mu$ to make ($\mathbf{P2.2}$) feasible and then we gradually increase the penalty parameter $\mu$ to force the slack variable $\phi$ approaching zero.
We note that ($\mathbf{P2.2}$) is equivalent to ($\mathbf{P2.1}$) when $\phi=0$.
In the following, we first transform the non-convex constraints \eqref{PF2_1b}, \eqref{PF2_1c}, and \eqref{PF2_2b} into convex constraints, and then we present an overall algorithm to solve the optimization problem ($\mathbf{P2.2}$).

\subsubsection{The transmission rate constraint \eqref{PF2_1b}}

The main challenge to tackle the non-convex constraint \eqref{PF2_1b} arises from the fact that the optimization variables in the constraint \eqref{PF2_1b} are coupled and the expression of the transmission rate given in \eqref{Rate2} is of a high complexity.
To overcome this challenge, we first introduce slack variables $\nu_k[n],\forall k,n$, and equivalently rewrite \eqref{PF2_1b} as
\begin{subequations}\label{PS5}
\begin{align}
&\frac{1}{N}\sum_{n=1}^N x_k[n]\nu_k[n]\geq \eta, ~\forall k,\label{PS5a}\\
&\log_2\left(1+\frac{\frac{\beta_0 P_k[n]}{\|\mathbf{q}_u[n]-\mathbf{w}_k\|^2+H^2}}{-\rho P_{u,\max}[n]\lambda_{u,u}\ln{\epsilon}+\sigma_u^2}\right)\geq v_k[n],~\forall k,n.\label{PS5b}
\end{align}
\end{subequations}
We note that, although \eqref{PS5a} and \eqref{PS5b} are still non-convex, they can be in more amenable forms than the original constraint \eqref{PF2_1b}. In the following, we focus on transforming \eqref{PS5a} and \eqref{PS5b} into convex constraints. For the non-convex constraint \eqref{PS5a}, we equivalently rewrite it as
\begin{align}\label{PS6}
\sum_{n=1}^N \left[(x_k[n]+\nu_k[n])^2-(x_k[n]-\nu_k[n])^2\right]\geq 4N\eta,~\forall k.
\end{align}
We note that \eqref{PS6} is in the form of the difference of two convex functions, which enables us to apply the first-order restrictive approximation to transform it into a convex constraint. As such, for given feasible points $\tilde{x}_k[n]$ and $\tilde{\nu}_k[n]$, $\forall k,n$, we rewrite \eqref{PS6} as
\begin{align}\label{PS7}
&\sum_{n=1}^N \big[2(\tilde{x}_k[n]+\tilde{\nu}_k[n])(x_k[n]-\tilde{x}_k[n]+\nu_k[n]-\tilde{\nu}_k[n])+\nonumber\\ &~~~~(\tilde{x}_k[n]+\tilde{\nu}_k[n])^2-(x_k[n]-\nu_k[n])^2\big]\geq 4N\eta,~\forall k.
\end{align}

For the non-convex constraint \eqref{PS5b}, we present the following lemma to aid tackling it.
\begin{lemma}\label{lemma2}
For given feasible points $\mathbf{\tilde{q}}_u[n]$ and $\tilde{P}_{u,\max}[n]$, $\forall n$, the following inequality
\begin{align}\label{lem2_1}
&\log_2\left(1\!+\!\frac{\frac{\beta_0P_k[n]}{\|\mathbf{q}_u[n]-\mathbf{w}_k\|^2+H^2}}{-\rho P_{u,\max}[n]\lambda_{u,u}\ln{\epsilon}\!+\!\sigma_u^2}\right)\nonumber\\
&~~\geq R^{lo}_k\left(\mathbf{q}_u[n],P_{u,\max}[n],\mathbf{\tilde{q}}_u[n],\tilde{P}_{u,\max}[n]\right),\forall k,n,
\end{align}
must hold, where $R^{lo}_k\left(\mathbf{q}_u[n],P_{u,\max}[n],\mathbf{\tilde{q}}_u[n],\tilde{P}_{u,\max}[n]\right)$ is defined in \eqref{lem2_2}, shown at the top of this page.
%\begin{align}\label{lem2_2}
%&R^{lo}_k\left(\mathbf{q}_u[n],P_{u,\max}[n],\mathbf{\tilde{q}}_u[n],\tilde{P}_{u,\max}[n]\right)\triangleq \log_2\left(1+\frac{\frac{\beta_0P_k[n]}{\|\mathbf{\tilde{q}}_u[n]-\mathbf{w}_k\|^2+H^2}}{-\rho \tilde{P}_{u,\max}[n]\lambda_{u,u}\ln{\epsilon}+\sigma_u^2}\right)+ \nonumber\\ &~~~~~~~~~~~~~~\frac{\frac{-\beta_0P_k[n]\left(\|\mathbf{q}_u[n]-\mathbf{w}_k\|^2-\|\mathbf{\tilde{q}}_u[n]-\mathbf{w}_k\|^2\right)}{\left(\|\mathbf{\tilde{q}}_u[n]-\mathbf{w}_k\|^2+H^2\right)\ln2}+\frac{\beta_0\rho\lambda_{u,u}\ln{\epsilon}P_k[n](P_{u,\max}[n]-\tilde{P}_{u,\max}[n])}{\left(-\rho \tilde{P}_{u,\max}[n]\lambda_{u,u}\ln{\epsilon}+\sigma_u^2\right)\ln2}}{\left(-\rho \tilde{P}_{u,\max}[n]\lambda_{u,u}\ln{\epsilon}+\sigma_u^2\right)\left(\|\mathbf{\tilde{q}}_u[n]-\mathbf{w}_k\|^2+H^2\right)+\beta_0P_k[n]}.
%\end{align}
%$R^{lo}_k\big(\mathbf{q}_u[n],P_{u,\max},\mathbf{\tilde{q}}_u[n],\tilde{P}_{u,\max}\big)$ is defined in \eqref{lem3_2}, which is shown at the top of this page.
\end{lemma}
\begin{IEEEproof}
The detailed proof is provided in Appendix \ref{App_lem2}.
\end{IEEEproof}
\setcounter{equation}{39}

We note that $R^{lo}_k\big(\mathbf{q}_u[n],P_{u,\max},\mathbf{\tilde{q}}_u[n],\tilde{P}_{u,\max}\big)$ is jointly concave with respect to the trajectory variable $\mathbf{q}_u[n]$ and the maximum AN transmit power $P_{u,\max}[n]$ for given feasible points $\mathbf{\tilde{q}}_u[n]$ and $\tilde{P}_{u,\max}[n]$. Following Lemma~\ref{lemma2}, the first-order restrictive approximation of \eqref{PS5b} is given by
\begin{align}\label{PS8}
R^{lo}_k\!\big(\mathbf{q}_u[n],P_{\!u,\max\!}[n],\mathbf{\tilde{q}}_u[n],\tilde{P}_{\!u,\max\!}[n]\big)\!\geq v_k[n],~\forall k,n.
\end{align}
We also note that \eqref{PS8} is a convex constraint due to the super-level set of a concave function.
So far, we have transformed the original non-convex ATR constraint \eqref{PF2_1b} into the convex constraints \eqref{PS7} and \eqref{PS8}.
%\newcounter{mytempeqncnt2}
%\begin{figure*}[tp]
%\normalsize
%\setcounter{mytempeqncnt2}{\value{equation}}
%\setcounter{equation}{40}
%\begin{align}\label{PS12}
%&\bar{\xi}_{k,m}^{up}\big(\mathbf{q}_u[n],P_{\!u,\max\!}[n],\mathbf{\tilde{q}}_u[n],\tilde{P}_{\!u,\max\!}[n]\big)\!\triangleq\!\left(\frac{\|\mathbf{q}_u[n]\!-\!\mathbf{w}_m\|^2+H^2}{\bar{\beta}_{k,m}[n]P_{u,\max}[n]}\!-\!\frac{\|\mathbf{\tilde{q}}_u[n]\!-\!\mathbf{w}_m\|^2+H^2}{\bar{\beta}_{k,m}[n]\tilde{P}_{u,\max}[n]}\right)\exp{\left(\frac{-\bar{\beta}_{k,m}[n]\tilde{P}_{u,\max}[n]}{\|\mathbf{\tilde{q}}_u[n]\!-\!\mathbf{w}_m\|^2\!+\!H^2}\right)}\times\nonumber\\
%&\underbrace{\left[\exp{\left(\frac{\bar{\beta}_{k,m}[n]\tilde{P}_{u,\max}[n]}{\|\mathbf{\tilde{q}}_u[n]\!-\!\mathbf{w}_m\|^2\!+\!H^2}\right)}\!-\!1\!-\!\frac{\bar{\beta}_{k,m}[n]\tilde{P}_{u,\max}[n]}{\|\mathbf{\tilde{q}}_u[n]\!-\!\mathbf{w}_m\|^2\!+\!H^2}\right]}_{\psi(\mathbf{\tilde{q}}_u[n],\tilde{P}_{u,\max}[n])}\!+\!\frac{\|\mathbf{\tilde{q}}_u[n]\!-\!\mathbf{w}_m\|^2\!+\!H^2}{\bar{\beta}_{k,m}[n]\tilde{P}_{u,\max}[n]}\left[1\!-\!\exp{\left(\frac{-\bar{\beta}_{k,m}[n]\tilde{P}_{u,\max}[n]}{\|\mathbf{\tilde{q}}_u[n]\!-\!\mathbf{w}_m\|^2\!+\!H^2}\right)}\right].
%\end{align}
%\setcounter{equation}{\value{mytempeqncnt2}}
%\hrulefill
%\vspace*{4pt}
%\end{figure*}

\subsubsection{Covertness constraint \eqref{PF2_1c}}

We first introduce slack variables $\omega_k[n],\forall k,n$, to convert the non-convex constraint \eqref{PF2_1c} into a more tractable form, which is given by
\begin{subequations}\label{PS9}
\begin{align}
&\sum_{k=1}^Kx_k[n]\omega_k[n]\geq 1-\varepsilon,~\forall n,\label{PS9a}\\
&1-\bar{\xi}_{k,m}^*[n]\geq\omega_k[n],~\forall k,m\in\mathcal{K}\setminus\{k\},\label{PS9b}
\end{align}
\end{subequations}
where $\bar{\xi}_{k,m}^*[n]$ is defined in \eqref{PS3}.
Similar to \eqref{PS5a}, for given feasible points $\tilde{x}_k[n]$ and $\tilde{\omega}_k[n]$, $\forall k,n$, we can rewrite \eqref{PS9a} as
\begin{align}\label{PS10}
&\sum_{k=1}^K\big[2(\tilde{x}_k[n]+\tilde{\omega}_k[n])(x_k[n]-\tilde{x}_k[n]+\omega_k[n]-\tilde{\omega}_k[n])+\nonumber\\
&(\tilde{x}_k[n]+\tilde{\omega}_k[n])^2-(x_k[n]-\omega_k[n])^2\big]\geq 4(1-\varepsilon),\forall n.
\end{align}
In the following, we focus on tackling the non-convex constraint \eqref{PS9b}. To this end, we first present the following lemma to determine the concavity of $\bar{\xi}_{k,m}^*[n]$.
\newcounter{mytempeqncnt2}
\begin{figure*}[tp]
\normalsize
\setcounter{mytempeqncnt2}{\value{equation}}
\setcounter{equation}{43}
\begin{align}\label{PS12}
&\bar{\xi}_{k,m}^{up}\big(\mathbf{q}_u[n],P_{\!u,\max\!}[n],\mathbf{\tilde{q}}_u[n],\tilde{P}_{\!u,\max\!}[n]\big)\!\triangleq\!\left(\frac{\|\mathbf{q}_u[n]\!-\!\mathbf{w}_m\|^2+H^2}{\bar{\beta}_{k,m}[n]P_{u,\max}[n]}\!-\!\frac{\|\mathbf{\tilde{q}}_u[n]\!-\!\mathbf{w}_m\|^2+H^2}{\bar{\beta}_{k,m}[n]\tilde{P}_{u,\max}[n]}\right)\exp{\left(\frac{-\bar{\beta}_{k,m}[n]\tilde{P}_{u,\max}[n]}{\|\mathbf{\tilde{q}}_u[n]\!-\!\mathbf{w}_m\|^2\!+\!H^2}\right)}\times\nonumber\\
&\underbrace{\left[\exp{\left(\frac{\bar{\beta}_{k,m}[n]\tilde{P}_{u,\max}[n]}{\|\mathbf{\tilde{q}}_u[n]\!-\!\mathbf{w}_m\|^2\!+\!H^2}\right)}\!-\!1\!-\!\frac{\bar{\beta}_{k,m}[n]\tilde{P}_{u,\max}[n]}{\|\mathbf{\tilde{q}}_u[n]\!-\!\mathbf{w}_m\|^2\!+\!H^2}\right]}_{\psi(\mathbf{\tilde{q}}_u[n],\tilde{P}_{u,\max}[n])}\!+\!\frac{\|\mathbf{\tilde{q}}_u[n]\!-\!\mathbf{w}_m\|^2\!+\!H^2}{\bar{\beta}_{k,m}[n]\tilde{P}_{u,\max}[n]}\left[1\!-\!\exp{\left(\frac{-\bar{\beta}_{k,m}[n]\tilde{P}_{u,\max}[n]}{\|\mathbf{\tilde{q}}_u[n]\!-\!\mathbf{w}_m\|^2\!+\!H^2}\right)}\right].
\end{align}
\setcounter{equation}{\value{mytempeqncnt2}}
\hrulefill
\vspace*{4pt}
\end{figure*}
\begin{lemma}\label{lemma3}
$\bar{\xi}_{k,m}^*[n]$ defined in \eqref{PS3} is a concave function of $\frac{\|\mathbf{q}_u[n]-\mathbf{w}_m\|^2+H^2\!}{P_{u,\max}[n]}$.
\end{lemma}
\begin{IEEEproof}
The result can be achieved by directly proving the second derivative of $\bar{\xi}_{k,m}^*[n]$ being negative. The detailed proof is omitted due to the space limitation.
\end{IEEEproof}

Following Lemma~\ref{lemma3}, the first-order restrictive approximation of \eqref{PS9b} is given by
\begin{align}\label{PS11}
&1-\bar{\xi}_{k,m}^{up}\big(\mathbf{q}_u[n],P_{\!u,\max\!}[n],\mathbf{\tilde{q}}_u[n],\tilde{P}_{\!u,\max\!}[n]\big)\geq\omega_k[n],\nonumber\\
&~~~~~~~~~~~~~~~~~~~~~~~~~~~~~~~~~~~~~~~\forall k,n,m\in\mathcal{K}\setminus\{k\},
\end{align}
where $\mathbf{\tilde{q}}_u[n]$ and $\tilde{P}_{u,\max}[n]$ are given feasible points. $\bar{\xi}_{k,m}^{up}\big(\mathbf{q}_u[n],P_{\!u,\max\!}[n],\mathbf{\tilde{q}}_u[n],\tilde{P}_{\!u,\max\!}[n]\big)$ is defined in \eqref{PS12}, shown at the top of this page.
%\begin{align}\label{PS12}
%&\bar{\xi}_{k,m}^{up}\big(\mathbf{q}_u[n],P_{\!u,\max\!}[n],\mathbf{\tilde{q}}_u[n],\tilde{P}_{\!u,\max\!}[n]\big)\!\triangleq\!\left(\frac{\|\mathbf{q}_u[n]\!-\!\mathbf{w}_m\|^2+H^2}{\bar{\beta}_{k,m}[n]P_{u,\max}[n]}\!-\!\frac{\|\mathbf{\tilde{q}}_u[n]\!-\!\mathbf{w}_m\|^2+H^2}{\bar{\beta}_{k,m}[n]\tilde{P}_{u,\max}[n]}\right)\times\nonumber\\
%&\exp{\left(\frac{-\bar{\beta}_{k,m}[n]\tilde{P}_{u,\max}[n]}{\|\mathbf{\tilde{q}}_u[n]\!-\!\mathbf{w}_m\|^2\!+\!H^2}\right)}
%\underbrace{\left[\exp{\left(\frac{\bar{\beta}_{k,m}[n]\tilde{P}_{u,\max}[n]}{\|\mathbf{\tilde{q}}_u[n]\!-\!\mathbf{w}_m\|^2\!+\!H^2}\right)}\!-\!1\!-\!\frac{\bar{\beta}_{k,m}[n]\tilde{P}_{u,\max}[n]}{\|\mathbf{\tilde{q}}_u[n]\!-\!\mathbf{w}_m\|^2\!+\!H^2}\right]}_{\psi(\mathbf{\tilde{q}}_u[n],\tilde{P}_{u,\max}[n])}+\nonumber\\
%&\frac{\|\mathbf{\tilde{q}}_u[n]\!-\!\mathbf{w}_m\|^2\!+\!H^2}{\bar{\beta}_{k,m}[n]\tilde{P}_{u,\max}[n]}\left[1\!-\!\exp{\left(\frac{-\bar{\beta}_{k,m}[n]\tilde{P}_{u,\max}[n]}{\|\mathbf{\tilde{q}}_u[n]\!-\!\mathbf{w}_m\|^2\!+\!H^2}\right)}\right].
%\end{align}
We also note that $\psi(\mathbf{\tilde{q}}_u[n],\tilde{P}_{u,\max}[n])$ defined in \eqref{PS12} must be greater than or equal to zero due to the fact that any convex function is lower bounded by its first-order approximation, i.e., $\exp{(x)}\geq 1+x$. In addition, $\frac{\|\mathbf{q}_u[n]-\mathbf{w}_m\|^2+H^2\!}{P_{u,\max}[n]}$ in \eqref{PS12} is a quadratic-over-linear function, which is a joint convex function with respect to the trajectory variable $\mathbf{q}_u[n]$ and the maximum AN transmit power $P_{u,\max}[n]$\cite{Boyd}. Following the above facts, we can conclude that $\bar{\xi}_{k,m}^{up}\big(\mathbf{q}_u[n],P_{\!u,\max\!}[n],\mathbf{\tilde{q}}_u[n],\tilde{P}_{\!u,\max\!}[n]\big)$ is jointly convex with respect to $\mathbf{q}_u[n]$ and $P_{u,\max}[n]$ for given feasible points $\mathbf{\tilde{q}}_u[n]$ and $\tilde{P}_{u,\max}[n]$, which implies that the constraint \eqref{PS11} is convex. So far, we have converted the non-convex constraint \eqref{PF2_1c} into the convex constraints \eqref{PS10} and \eqref{PS11}.

\setcounter{equation}{44}
\subsubsection{The constraint \eqref{PF2_2b}}
We note that the left hand side (LHS) of the constraint \eqref{PF2_2b} is in the form of a sum of a linear function and a convex function, while the right hand side (RHS) of \eqref{PF2_2b} is a linear function. This special form allows us to apply the first-order restrictive approximation to transform it into a linear constraint, which is given by
\begin{align}\label{PS4}
\sum_{n=1}^N\sum_{k=1}^K\left(x_k[n]+\tilde{x}_k[n]^2-2\tilde{x}_k[n]x_k[n]\right)\leq \phi,
\end{align}
where $\tilde{x}_k[n]$, $\forall k,n$, are given feasible points. We note that the constraint \eqref{PS4}
is stricter than the original constraint \eqref{PF2_2b} due to that any convex function is lower bounded by its first-order approximation, which leads to that any feasible solution to \eqref{PS4} is also feasible to \eqref{PF2_2b}.

Following the above transformations detailed in 1), 2), and 3), we rewrite the optimization problem ($\mathbf{P2.2}$) as
\begin{align}
&(\mathbf{P2.3}):~
\max_{\eta,\mathbf{Q},\mathbf{X},\mathbf{P}_{\mathrm{U}},\mathbf{V},\mathbf{W},\phi}~\eta-\mu\phi\nonumber\\
&\mathrm{s.t.}~\eqref{PS7},\eqref{PS8},\eqref{PS10},\eqref{PS11}, \eqref{PS4},\nonumber\\
&~~~~~\eqref{PF1c},\eqref{PF1e},\eqref{PF1f},\eqref{PF1g}, \eqref{PS2b}, \notag
\end{align}
where $\mathbf{V}\triangleq\{\nu_k[n],\forall k,n\}$ and $\mathbf{W}\triangleq\{\omega_k[n],\forall k,n\}$.

We note that the optimization problem ($\mathbf{P2.3}$) is with a linear objective function and a convex constraint set. As such, it is a standard convex optimization problem. For given penalty parameter $\mu$ and feasible points $(\tilde{x}_k[n],\mathbf{\tilde{q}}_u[n], \tilde{P}_{u,\max}[n],\tilde{\nu}_k[n],\tilde{\omega}_k[n])$, $\forall k,n$, ($\mathbf{P2.3}$) can be efficiently solved by convex optimization solver such as CVX~\cite{Boyd}. We note that the constraint set of ($\mathbf{P2.3}$) is stricter than that of ($\mathbf{P2.2}$), since the first-order restrictive approximation was applied to transform ($\mathbf{P2.2}$) into ($\mathbf{P2.3}$). As such, the optimal solution to the optimization problem ($\mathbf{P2.3}$) is also feasible to the optimization problem ($\mathbf{P2.2}$).

\begin{algorithm}[t]
\caption{P-SCA scheme for Solving Problem ($\mathbf{P1}$)}\label{alg1}
\begin{algorithmic}[1]
\STATE Given feasible points $(\tilde{x}_k^0[n], \mathbf{\tilde{q}}_u^0[n], \!\tilde{P}_{\!u,\max\!}^0[n]\!,\!\tilde{\nu}_k^0[n]\!,\tilde{\omega}_k^0[n])$, $\forall k,n$, and an initial penalty parameter $\mu^0$; $r=0$.
\REPEAT
\STATE {Solve ($\mathbf{P2.3}$) with given feasible points $(\tilde{x}_k^r[n], \mathbf{\tilde{q}}_u^r[n], \tilde{P}_{u,\max}^r[n],\tilde{\nu}_k^r[n],\tilde{\omega}_k^r[n])$, $\forall k,n$, and obtain $(x_k^{r+1}[n], \mathbf{q}_u^{r+1}[n], P_{u,\max}^{r+1}[n],\nu_k^{r+1}[n],\omega_k^{r+1}[n])$.}
\STATE {Set  $(\tilde{x}_k^r[n], \mathbf{\tilde{q}}_u^r[n], \tilde{P}_{\!u,\max\!}^r[n],\tilde{\nu}_k^r[n],\tilde{\omega}_k^r[n])=(x_k^{r\!+\!1}[n],\mathbf{q}_u^{r\!+\!1}[n], P_{u,\max}^{r\!+\!1}[n],\nu_k^{r\!+\!1}[n],\omega_k^{r\!+\!1}[n])$, and update $\mu^{r+1}=\min\{c\mu^r,\mu_{\max}\}$; $r=r\!+\!1$.}
\UNTIL {Convergence.}
\end{algorithmic}
\end{algorithm}

In the following, we present an overall P-SCA scheme to solve ($\mathbf{P2.2}$). We first note that ($\mathbf{P2.2}$) is equivalent to ($\mathbf{P1}$) when $\phi\rightarrow 0$. Following the principle of P-SCA, we solve ($\mathbf{P2.2}$) by successively solving ($\mathbf{P2.3}$) for given feasible points $(\tilde{x}_k^r[n], \mathbf{\tilde{q}}_u^r[n], \tilde{P}_{\!u,\max\!}^r[n],\tilde{\nu}_k^r[n],\tilde{\omega}_k^r[n])$, $\forall k, n$, and  penalty parameter $\mu$, where $r$ denotes the $r$-th iteration. The detailed algorithm is given in Algorithm~\ref{alg1}.
%We observe that the feasible point $(\tilde{x}_k^{i}[n], \tilde{P}_{\!u,\max\!}^i[n],\mathbf{\tilde{q}}_u^i[n])$ and penalty parameter $\mu$ are updated at each iteration.
We note that the penalty parameter $\mu$ determines the relaxation level of ($\mathbf{P2.2}$). In general, a large $\mu$ strongly forces $\phi\rightarrow 0$, resulting in $x_k[n]\in\{0,1\}$. Algorithm \ref{alg1} starts with a small value of $\mu$ to provide a larger feasible set for $x_k[n]$, and then increases the penalty parameter $\mu$ with a constant $c>1$ at each iteration until a large upper bound $\mu_{\max}$ is achieved to guarantee that $\phi=0$.
Additionally, Algorithm \ref{alg1} is guaranteed to converge to a stationary point, i.e., fulfilling the Karush-Kuhn-Tucker (KKT) optimality conditions of ($\mathbf{P2.1}$). The proof is similar to that given in \cite{Vu2016Max} and thus is omitted here for brevity.

We note that in Algorithm~\ref{alg1}, it is very critical to provide an initial feasible UAV trajectory, since it not only determines the feasibility of ($\mathbf{P2.3}$) and the convergence speed of Algorithm~\ref{alg1}, but it directly affects the convergence solution to Algorithm~\ref{alg1}~\cite{JointWu2018}.
Thus, in the next subsection, we propose an efficient trajectory initialization scheme.

\subsection{Trajectory Initialization Scheme}

The SHAF (i.e., Successive Hover-and-Fly) trajectory is beneficial to the uplink transmission, since it improves the communication channel quality from the ground users to the UAV. As such, it is reasonable for the UAV to fly with the SHAF trajectory, i.e., the UAV sequentially hovers at each of ground users for a certain time and flies from one user to another with the maximum speed $V_{\max}$ when the flight period $T$ is sufficiently large for the UAV to reach the top of each ground user.

In the following, we first determine the UAV's traveling path to visit all the $K$ ground users with the minimum flying distance so as to
minimize the total flying time and we denote the minimum flying time as $T_{\min}$. Then, the
remaining time $T-T_{\min}$ is reasonably allocated by the UAV to hover above the ground users. We note that the flying distance minimization problem is a
classic traveling salesman problem (TSP) and its optimal solution usually occurs a high computational complexity~\cite{LaporteThe}. To maintain a low complexity, here we apply the nearest-neighbour algorithm with complexity $\mathcal{O}(K^3)$  to determine an
approximation solution to the TSP. The detailed nearest-neighbour algorithm is given in \cite{LaporteThe} and omitted here for brevity.

Intuitively, if the minimum distance from the $k$-th user to other users is $d_k$ and the $m$-th ($m\neq k$) user to other users is $d_m$, the UAV should hover more time over the $k$-th user when $d_k<d_m$ in order to maximize the minimum ATR among all the ground users. This is due to the fact that, the nearest detector to the $k$-th SU is relatively stronger than the nearest detector to the $m$-th SU, which leads to a smaller ATR at the $k$-th SU. Following this fact, the total hovering time $T-T_{\min}$ can be proportionally allocated among the ground users as
$\tilde{T}_k=\frac{\frac{1}{d_k}(T-T_{\min})}{\sum_{m=1}^K\frac{1}{d_m}}$, $\forall k$,
where $\tilde{T}_k$ is the hovering time over user $k$.

When $T<T_{\min}$, the UAV's trajectory based on nearest-neighbour algorithm is infeasible, since flying time is not sufficient for the UAV to reach the upright locations of all the ground users.
To determine the initial UAV's trajectory in this case, we first find the geometric center of the ground users, which is given by $\mathbf{w}_0=\frac{\sum_{k=1}^K\mathbf{w}_k}{K}$. Then, we reconstruct the trajectory by down-scaling the SHAF trajectory $\{\mathbf{\hat{q}}(t)$, $0\leq t\leq T_{\min}\}$ (which is obtained by applying the nearest-neighbour algorithm) linearly towards the geometric center, such that total flying distance exactly equals to $V_{\max}T$, i.e.,
\begin{align}\label{Sub1_1}
\mathbf{q}(t)=\mathbf{\hat{q}}(t/{\varpi})+(1-\varpi)\left(\mathbf{w}_0-\mathbf{\hat{q}}(t/{\varpi})\right),
\end{align}
where $\varpi=\frac{T}{T_{\min}}$ denotes the scaling factor. Following the above method, we obtain a continuous UAV's trajectory, which can be discredited into the sequence $\{\mathbf{q}[n]\}_{n=1}^N$.

\section{numerical results}

In this section, we provide numerical results to evaluate the performance of the developed UAV-enabled covert data collection based on P-SCA. To demonstrate the benefit of our developed design, we compare our P-SCA scheme with a benchmark (BM) scheme, where in the latter scheme the UAV's trajectory is fixed as the SHAF trajectory (detailed in Section IV-C) and only the UAV's maximum AN transmit power and user scheduling strategy are optimized.
Without other statements, the system parameters are set as: $P_{\max}^u=36~\mathrm{dBm}$, $P_k[n]=30~\mathrm{dBm}, \forall k,n$, $V_{\max}=6~\mathrm{m/s}$, $H=100~\mathrm{m}$, $\delta_t=1~\mathrm{s}$, $\lambda_{u,u}=-60~\mathrm{dB}$, $\rho=-60~\mathrm{dB}$, $\sigma_u^2=\sigma_m^2=-110~\mathrm{dBm}$, $\beta_0=-60~\mathrm{dB}$, $T=240~\mathrm{s}$, $\epsilon=0.05$, and $\varepsilon=0.03$. In addition, we consider that there are four users on the ground, i.e., $K=4$, whose horizontal coordinates are set as $[200, 0]^T$, $[0, 120]^T$, $[-200, 0]^T$, and $[0, -120]^T$, respectively.
\begin{figure}[!t]
  \centering
  % Requires \usepackage{graphicx}
  \includegraphics[width=3.49in, height=2.8in]{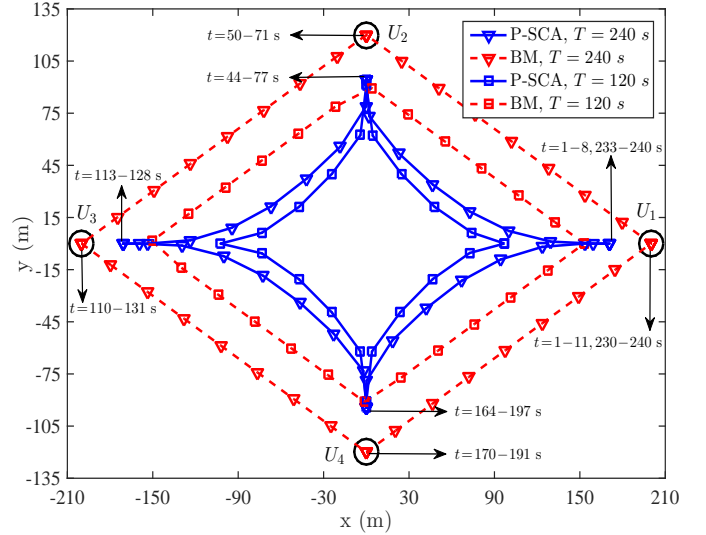}
  \caption{UAV's trajectories achieved by the P-SCA amd BM schemes for different values of the flight period $T$.}\label{Trajectory}
\end{figure}

\begin{figure}[!t]
  \centering
  % Requires \usepackage{graphicx}
  \includegraphics[width=3.49in, height=2.8in]{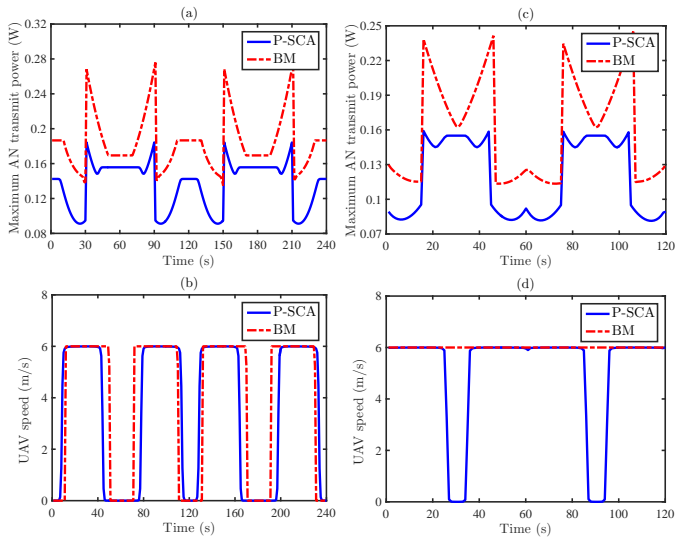}
  \caption{The UAV's transmit power and speed for different values of the flight period $T$, where $T=240~\mathrm{s}$ for (a) and (b), while $T=120~\mathrm{s}$ for (c) and (d).}\label{Pow_Speed}
\end{figure}

In Fig.~\ref{Trajectory}, we plot the trajectories of the UAV achieved by our P-SCA scheme and the BM scheme with different values of the flight period $T$, where the $k$-th ground user is denoted as $U_k$ ($k\in \mathcal{K}$) and the location of each user is marked with $\bigcirc$.
%In this figure, as expected we first observe that the UAV's trajectory achieved by the BMS scheme is consistent with our designed SHAF trajectory, where $T=240~\mathrm{s}$ and $T=120~\mathrm{s}$ corresponding to $T$ is sufficiently large so that the UAV is able to reach the top of each ground user and the flying time is not sufficient for the UAV to visit all the ground users, respectively.
In this figure, we first observe that the trajectory achieved by the P-SCA scheme always shrinks inward relative to that achieved by the BM scheme. This is due to the fact that, when the UAV's trajectory is closer to the USUs, the UAV's maximum AN transmit power can be relatively smaller to reduce self-interference and improve the max-min ATR (i.e., average transmission rate) while maintaining a certain level of covertness.
This is confirmed by Fig.~\ref{Pow_Speed}(a) and Fig.~\ref{Pow_Speed}(c), where the UAV's maximum AN transmit power in the P-SCA scheme is significantly smaller than that in the BM scheme.
In Fig.~\ref{Trajectory}, we also observe that each hovering location achieved by the P-SCA scheme is close to but slightly away from the upright location of each user, even when the flight period is sufficient (e.g., $T=240~\mathrm{s}$). We note that these hovering locations are generally to strike a tradeoff between the communication performance from the SU (i.e., scheduled user) to the UAV and the detection performance at the USUs (i.e., unscheduled users).

In Fig.~\ref{Trajectory}, Fig.~\ref{Pow_Speed}(b), and Fig.~\ref{Pow_Speed}(d), we observe that the UAV's hovering time around $U_2$ and $U_4$ is larger than that around $U_1$ and $U_3$ in our P-SCA scheme, since the distance between $U_2$ and $U_4$ is smaller than that between $U_1$ and $U_3$.
This is the reason why the UAV's maximum AN transmit power is larger when $U_2$ or $U_4$ is scheduled than that when $U_1$ or $U_3$ is scheduled, shown in Fig.~\ref{Pow_Speed}(a) and Fig.~\ref{Pow_Speed}(c). This can improve the max-min ATR among all the ground users, since the ATR decreases as the maximum AN transmit power increases due to the self-interference.

\begin{figure}[!t]
  \centering
  % Requires \usepackage{graphicx}
  \includegraphics[width=3.49in, height=2.8in]{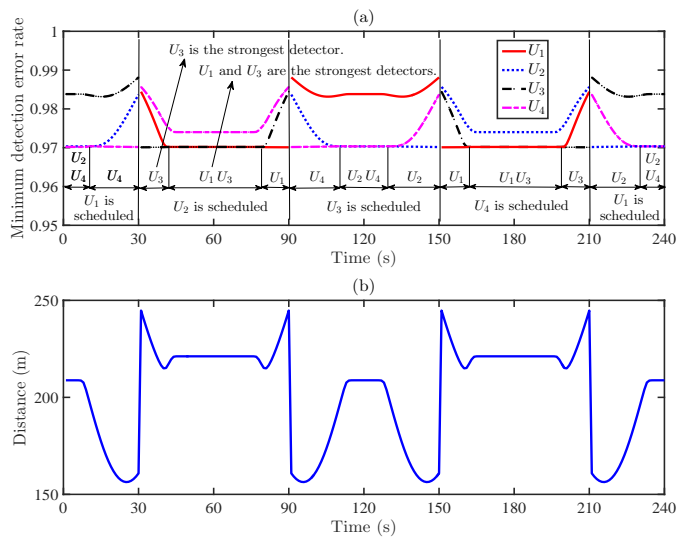}
  \caption{Each USU's minimum detection error rate and the distance from the UAV to the strongest detector achieved by the P-SCA scheme versus time, where the flight period is set as $T=240~\mathrm{s}$ and the covertness level parameter is set as $\varepsilon=0.03$.}\label{Detection_Distance}
\end{figure}

In Fig.~\ref{Detection_Distance}(a), we plot each USU's minimum detection error rate achieved by the P-SCA scheme versus time. In this figure, we first observe that the minimum detection error rate of the strongest detector in the USUs is equal to the required covertness (i.e., $\xi_{k,m}^\ast[n]= 1-\varepsilon$). This is due to the fact that the max-min ATR monotonically decreases with the UAV's maximum AN transmit power while the minimum detection error rate $\xi_{k,m}^\ast[n]$ at each USU is a monotonically increasing function of the maximum AN transmit power.
In the following, we take the UAV flight time from $30~\mathrm{s}$ to $90~\mathrm{s}$ as an example to draw some insights on the system, where $U_2$ is scheduled in this period of time.
We observe that $U_3$ is the strongest detector at the initial stage of this period, since the UAV is farthest from $U_3$ (i.e., $U_3$ receives the least AN interference).
Then, $U_1$ and $U_3$ are the strongest detectors. This is due to the fact that the UAV hovers around $U_2$ at this period of time, while the distance from the UAV to $U_1$ is the same as that from the UAV to $U_3$, which is larger than the distance from the UAV to $U_4$.
Finally, $U_1$ becomes the strongest detector for reasons similar to the initial stage. In Fig.~\ref{Detection_Distance}(b), we plot the distance from the UAV to the strongest detector achieved by the P-SCA scheme versus time. In this figure, we observe that the curve of the UAV's maximum AN transmit power achieved by the P-SCA scheme shown in Fig.~\ref{Pow_Speed}(a) has the same trend as the curve shown in Fig.~\ref{Detection_Distance}(b), which indicates that the UAV's maximum AN transmit power achieved by P-SCA scheme
is dominated by the distance from the UAV to the strongest detector. This is also the reason why the maximum AN transmit power curve changes according to Fig.~\ref{Pow_Speed}(a).
\begin{figure}[!t]
  \centering
  % Requires \usepackage{graphicx}
  \includegraphics[width=3.49in, height=2.8in]{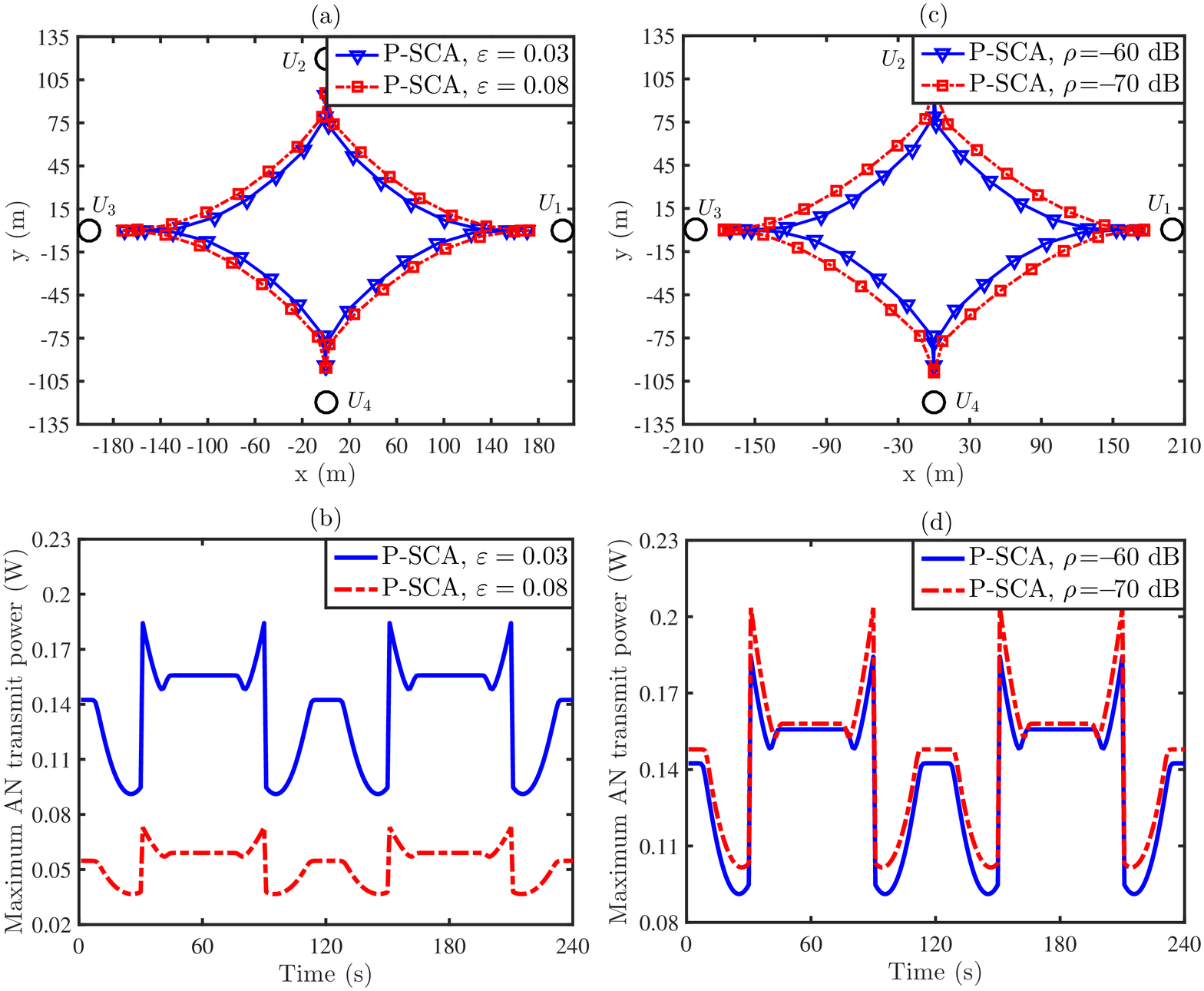}
  \caption{UAV's trajectories and the maximum AN transmit power achieved by P-SCA scheme for different values of $\varepsilon$ and $\rho$.}\label{Traj_rho_w}
\end{figure}

In Fig.~\ref{Traj_rho_w}, we plot the UAV's trajectories and the corresponding maximum AN transmit power achieved by the P-SCA scheme for different levels of required covertness (i.e., different values of $\varepsilon$) and different self-interference levels $\rho$. In Fig.~\ref{Traj_rho_w}(a) and Fig.~\ref{Traj_rho_w}(b), we observe that, as $\varepsilon$ increases, the UAV's trajectory expands outward and its maximum AN transmit power decreases. This is due to the fact that the covertness constraint (i.e., $\xi_{k,m}^\ast[n]\geq 1-\varepsilon$) becomes stricter as $\varepsilon$ decreases. Then, the UAV prefers to select a trajectory closer to each user to improve the channel quality and use a smaller maximum AN transmit power to reduce the self-interference while satisfying the covertness constraint.
In Fig.~\ref{Traj_rho_w}(c) and Fig.~\ref{Traj_rho_w}(d), as expected we observe that UAV's trajectory expands outward as $\rho$ decreases, while the maximum AN transmit power increases as $\rho$ decreases.
%This is due to the fact that the UAV's trajectory has a larger impact on the considered covert communications.

\begin{figure}[!t]
  \centering
  % Requires \usepackage{graphicx}
  \includegraphics[width=3.49in, height=2.8in]{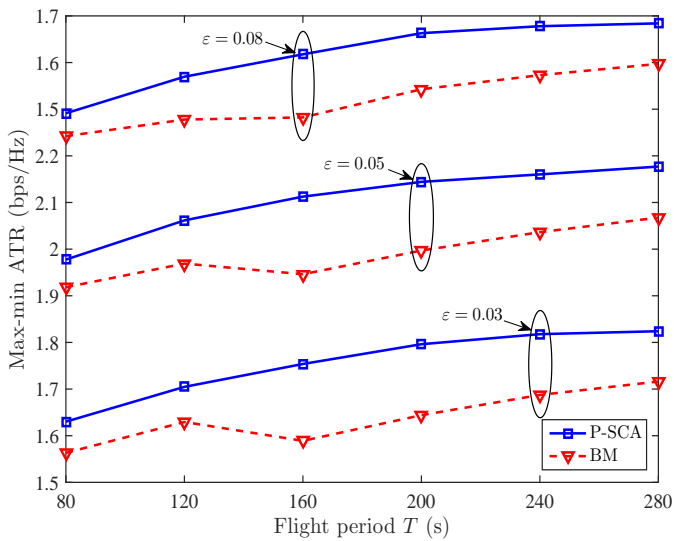}
  \caption{Max-min ATR (i.e., average transmission rate) achieved
by the P-SCA and BM schemes versus the flight period $T$ for different values of $\varepsilon$.}\label{Rate_T_rhow}
\end{figure}

In Fig.~\ref{Rate_T_rhow}, we plot the max-min ATRs achieved
by the P-SCA and BM schemes versus the flight period $T$ for different values of $\varepsilon$. In this figure, we first observe that the achieved max-min ATR by the P-SCA scheme monotonically increases with $T$. This is due to the fact that a large flight period $T$ offers a larger degree of freedom for us to design the UAV's covert data collection.
However, we also observe that the max-min ATR achieved by the BM scheme is not monotonically increasing with respect to $T$. This is mainly due to the fact that the UAV's trajectory is hardly determined, not softly designed as in the P-SCA scheme. This observation demonstrates the necessity of designing the UAV's trajectory in the considered covert data collection system (as we did in the P-SCA scheme). Furthermore, in this figure we observe that
the proposed P-SCA scheme always achieves a higher max-min ATR than the BM scheme, which demonstrates the advantages of the joint design of the UAV's trajectory, maximum AN transmit power, and the user scheduling strategy. Finally, as expected we observe that the max-min ATR significantly increases as the $\varepsilon$ increases, since a smaller $\varepsilon$ leads to a stricter covertness constraint, which is the key performance limiting factor in covert communications.

%\begin{figure}[!t]
%  \centering
%  % Requires \usepackage{graphicx}
%  \includegraphics[width=2.9in, height=2.3in]{Rate_Pk4.eps}
%  \caption{Max-min ATR achieved
%by the P-SCA and BMS scheme versus the transmit power of each user for different self-interference levels $\varepsilon$.}\label{Rate_Pk}
%\end{figure}
%
%In Fig.~\ref{Rate_Pk}, we plot max-min ATR achieved
%by the P-SCA and BMS scheme versus the transmit power of each user for different self-interference levels $\rho$. In this figure, as expected we observe that the max-min ATR increases as the self-interference $\rho$ decreases. We also observe that the max-min ATR approaches to a fixed value as the transmit power of SU increases. This is due to the fact that although a larger transmit power at the SU will benefit to the uplink transmission, it is also easier for the USUs to detect its transmission. To maintain the same level of covertness requirement, UAV need to increase its maximum AN transmit power. As such, the benefits of increased the $P_k[n]$ are gradually offset by the self-interference, which makes the max-min ATR approaches to a fixed value as the $P_k[n]$ increases.

%=======================================================================================
\section{Conclusion}
In this work, for the first time, we considered covert data collection in UAV wireless networks based on covert communication techniques.
We first determined the covertness constraint explicitly by analyzing the detection performance at each USU in terms of deriving the minimum detection error rate. Then, we formulated an optimization problem to maximize the minimum ATR among all ground users to the UAV subject to the covertness constraint and other practical constraints, e.g., the UAV's mobility constraints.
To tackle this mixed-integer optimization problem, a novel P-SCA scheme is developed to design the UAV's trajectory and the maximum AN transmit power as well as the user scheduling strategy. Our examinations showed that the developed P-SCA scheme always outperforms a benchmark scheme. Interestingly, we found that, as the covertness level increases, the UAV's trajectory achieved by the P-SCA schemes always shrinks inward to the centre determined by the locations of all the ground users and the UAV's maximum AN transmit power increases.

\appendices
%========================================================================================

\section{Proof of Lemma \ref{lemma2}}\label{App_lem2}
To prove Lemma~\ref{lemma2}, we first determine the convexity of $f(x_1,x_2)$ with respect to $x_1$ and $x_2$, where $f(x_1,x_2)$ is defined as
\begin{align}\label{App_lem2_1}
f(x_1,x_2)=\log_2\left(1+\frac{a}{x_1x_2}\right),
\end{align}
while $a\geq 0$, $x_1>0$ and $x_2>0$. To this end, we present the Hessian matrix of $f(x_1,x_2)$, which is given by
\begin{align}\label{App_lem2_2}
&\triangledown^2f(x_1,x_2)
%\frac{c}{(c+x_1x_2)^2\ln{2}}
%\left[\begin{matrix}
%   \frac{c}{x_1^2}+\frac{2x_2}{x_1} & 1  \\
%   1 & \frac{c}{x_2^2}+\frac{2x_1}{x_2}
%\end{matrix}\right]\nonumber\\
=\frac{a}{(a+x_1x_2)^2\ln{2}}\left[
\begin{matrix}
   \frac{a}{x_1^2}+\frac{x_2}{x_1} & 0  \\
   0 & \frac{a}{x_2^2}+\frac{x_1}{x_2}
\end{matrix}\right]\nonumber\\
&~~~~~~~~+\frac{a}{x_1x_2(a+x_1x_2)^2\ln{2}}
\left[\begin{matrix}
   x_2 \\
   x_1
\end{matrix}\right]
\left[\begin{matrix}
   x_2 & x_1
\end{matrix}\right]\succeq \mathbf{0}.
\end{align}
Following \eqref{App_lem2_2}, we can conclude that $f(x_1,x_2)$ is jointly convex with respect to $x_1$ and $x_2$. We note that any convex function is lower bounded by its first-order approximation~\cite{Boyd}. As such, for given feasible points $\tilde{x}_1$ and $\tilde{x}_2$, we have
\begin{align}\label{App_lem2_3}
&f(x_1,x_2)\geq\nonumber\\
&\log_2\left(1+\frac{a}{\tilde{x}_1\tilde{x}_2}\right)+\frac{-a(x_1-\tilde{x}_1)}{\tilde{x}_1(\tilde{x}_1\tilde{x}_2+a)}+\frac{-a(x_2-\tilde{x}_2)}{\tilde{x}_2(\tilde{x}_1\tilde{x}_2+a)}.
\end{align}
We note that $\log_2\Big(1+\frac{\frac{\beta_0P_k[n]}{\|\mathbf{q}_u[n]-\mathbf{w}_k\|^2+H^2}}{-\rho P_{u,\max}[n]\lambda_{u,u}\ln{\epsilon}+\sigma_u^2}\Big)
$ in \eqref{lem2_1} is in a similar form as $f(x_1,x_2)$. In addition, we note that $\beta_0P_k[n]\geq 0$, $\|\mathbf{q}_u[n]-\mathbf{w}_k\|^2+H^2>0$ and $-\rho P_{u,\max}[n]\lambda_{u,u}\ln{\epsilon}+\sigma_u^2>0$ always hold.
Following these facts, by replacing $a$, $x_1$, $x_2$, $\tilde{x}_1$, and $\tilde{x}_2$ in \eqref{App_lem2_3} with $\beta_0P_k[n]$, $\|\mathbf{q}_u[n]-\mathbf{w}_k\|^2+H^2$, $-\rho P_{u,\max}[n]\lambda_{u,u}\ln{\epsilon}+\sigma_u^2$, $\|\mathbf{\tilde{q}}_u[n]-\mathbf{w}_k\|^2+H^2$, and $-\rho \tilde{P}_{u,\max}[n]\lambda_{u,u}\ln{\epsilon}+\sigma_u^2$, respectively, we obtain the result given in \eqref{lem2_1}. This completes the proof of Lemma~\ref{lemma2}.

\bibliographystyle{IEEEtran}
\bibliography{IEEEfull,UAV_Covert_Upplink}

% Note that IEEE does not put floats in the very first column - or typically
% anywhere on the first page for that matter. Also, in-text middle ("here")
% positioning is not used. Most IEEE journals use top floats exclusively.
% Note that, LaTeX2e, unlike IEEE journals, places footnotes above bottom
% floats. This can be corrected via the \fnbelowfloat command of the
% stfloats package.

\ifCLASSOPTIONcaptionsoff
  \newpage
\fi

%\begin{thebibliography}{1}

%XXXXXXXX

%\end{thebibliography}
%\clearpage
%\newpage

\end{document}